\documentclass[twocolumn,aps,prd,superscriptaddress,amsmath,amssymb,nofootinbib,nobibnotes]{revtex4-2}
\usepackage{anyfontsize}
\usepackage{amssymb}
\usepackage{amsmath}
\usepackage{amsfonts}
\usepackage{amsbsy}
\usepackage[utf8]{inputenc} 
\usepackage{newtxtext}
\usepackage{newtxmath}

\usepackage[normalem]{ulem}

\usepackage{tabularray}
\UseTblrLibrary{diagbox}

\usepackage{tensor}
\usepackage{mathrsfs}
\usepackage{slashed}

\usepackage{bm}

\usepackage[utf8]{inputenc}

\usepackage[dvipsnames]{xcolor}
\usepackage{verbatim,mathtools,needspace,enumitem,etoolbox}
\usepackage{physics}

\definecolor{linkcolor}{rgb}{0.0,0.3,0.5}

\usepackage[unicode, colorlinks=true, linkcolor=linkcolor, citecolor=linkcolor, filecolor=linkcolor,urlcolor=linkcolor, pdfusetitle]{hyperref}

\usepackage{epstopdf}

\usepackage{bm}
\usepackage{dcolumn}

\usepackage{longtable}
\usepackage{enumerate}
\usepackage[normalem]{ulem}

\hypersetup{colorlinks=true,
citecolor=blue,
linkcolor=blue,
urlcolor=blue}
\usepackage{subfigure}
\usepackage{float}
\usepackage{mathrsfs}
\usepackage[dvipsnames]{xcolor}
\usepackage{soul}

\definecolor{darkgreen}{RGB}{1,212,57}

\usepackage{enumitem}

\usepackage{graphicx}
\usepackage{dcolumn}
\usepackage{bm}

\begin{document}

\title{Spinning generalizations of Majumdar-Papapetrou multi-black hole spacetimes: \\ light rings, lensing and shadows}

\author{Zeus S. Moreira}
\email{zeus.moreira@icen.ufpa.br}
\affiliation{%
	Programa de Pós-Graduação em Física, Universidade Federal do Pará, 66075-110, Belém, Par{\'a}, Brazil 
}%
\author{Carlos A. R. Herdeiro}
\email{herdeiro@ua.pt}
\affiliation{%
	Programa de Pós-Graduação em Física, Universidade Federal do Pará, 66075-110, Belém, Par{\'a}, Brazil 
}%
\affiliation{Departamento de Matem\'atica da Universidade de Aveiro and Center for Research and Development  in Mathematics and Applications (CIDMA), Campus de Santiago, 3810-193 Aveiro, Portugal.}
\author{Luís C. B. Crispino}
\email{crispino@ufpa.br}
\affiliation{%
	Programa de Pós-Graduação em Física, Universidade Federal do Pará, 66075-110, Belém, Par{\'a}, Brazil 
}%


\begin{abstract}
	A generalization of the Majumdar-Papapetrou multi-black hole spacetime was recently constructed by Teo and Wan~\cite{Teo:2023wfd}, describing charged and spinning (extremal) balanced black holes in asymptotically flat spacetime. We explore the dynamics of null geodesics on this geometry, focusing on the two-center solution. Using the topological charge formalism, we show that various light ring arrangements arise from different choices of individual angular momenta: light rings with opposite topological charges can merge and annihilate each other, resulting in configurations with a total of 4, 6, or 8 light rings. Using backward ray-tracing, we obtained the shadow and lensing of these spacetimes. The former, in particular, closely resembles those for the double-Kerr metric. 
\end{abstract}

\maketitle



\section{Introduction}

The recent observations by LIGO, Virgo, and KAGRA have established the study of gravitational waves as a central area of research in theoretical physics~\cite{LIGOScientific:2016aoc,KAGRA:2021vkt,LIGOScientific:2024elc}. Dynamical binary black hole (BH) solutions can model the complex dynamics of common gravitational wave sources. Yet, constructing these solutions in General Relativity (GR) is exceptionally challenging, as the problem lacks generic symmetries and involves fundamentally time-dependent processes. 

An intermediate simpler way, albeit limited in scope, to get a glimpse on how BHs interact with each other is to analyze exact stationary solutions involving multiple BHs. In such solutions the BHs must be ``fixed'', to equilibrium to be achieved. This can be done by either fine tuning their physical parameters (distances, masses, angular momenta, gauge charges, external fields,...) or by  introducing  artificial structures to hold the BHs in place, such as conical singularities.

In four dimensional vacuum GR, asymptotically flat multi-centered BH configurations cannot remain in equilibrium without the presence of naked singularities~\cite{bunting1987nonexistence,Costa:2009wj,Neugebauer:2011qb,Hennig:2011fp}. For example, the double Schwarzschild solution~\cite{bach1922neue} and the double Kerr solution~\cite{kramer1980superposition} require conical singularities. Even without resorting to gauge fields, however, conical singularities can be removed either by introducing additional (non-gauge) fields, such as scalar fields~\cite{Herdeiro:2023mpt,Herdeiro:2023roz}, or by waiving asymptotic flatness~\cite{Astorino:2021dju,Dias:2023rde}.

The introduction of gauge fields, on the other hand, opens up new possibilities. In Newtonian mechanics, the equilibrium between two charged, massive particles is achieved when the product of their masses equals the product of their electric charges (in appropriate units). The inherent non-linearities of GR would seem to preclude a straightforward generalization of this simple rule. Remarkably, however, in Einstein-Maxwell theory an equally simple (and related) rule applies: two (or more) BH configurations are possible if the BHs have the same (unitary) charge to mass ratio. Such spacetimes are constructed by superimposing two (or more) extremal Reissner–Nordström (RN) BHs, leading to the Majumdar-Papapetrou (MP) solution~\cite{Majumdar:1947eu,Papaetrou:1947ib}.

There have been attempts to extend the static MP solution to rotating spacetimes within Einstein-Maxwell theory~\cite{Israel:1972vx,Perjes:1971gv}. It turns out that the obtained generalizations describe the interaction between naked singularities rather than BHs~\cite{Hartle:1972ya}. On the other hand, enlarging the model from Einstein-Maxwell to Einstein-Maxwell-dilaton with the Kaluza-Klein (KK) coupling between the dilaton and Maxwell field - hereafter dubbed \textit{KK theory} -, Teo and Wan constructed a generalized version of the MP solution that accommodate rotating, electrically and magnetically charged BHs~\cite{Teo:2023wfd} that reduce to the standard MP of Einstein-Maxwell in the static limit. In other words, the dilaton is sourced by rotation.  Moreover, the Teo-Wan (TW) spacetime represents a regular superposition of the well-known rotating, charged BHs in KK theory, the Rasheed-Larsen (RL) BHs~\cite{Rasheed:1995zv,Larsen:1999pp}.

The TW metric is much simpler than the double-Kerr solution, making it a convenient laboratory for rotational effects on physical observables. Additionally, it is free of conical singularities, which makes it more physically appealing. Even though it should not be faced as a representation of (astro)physical reality, as it remains inherently dyonic and with extremal horizons,  its theoretical advantages, motivate us to analyze its null geodesic flow, exploring the effects of rotation in a multi-BH system.

A key motivation for studying null geodesics on BH spacestimes has been the Event Horizon Telescope's groundbreaking imaging of the supermassive BHs M87* and SgrA*~\cite{EventHorizonTelescope:2019dse,EventHorizonTelescope:2022wkp}. This milestone has fueled significant interest in investigating the behavior of light near BHs, as such trajectories offer critical insights into the structure and properties of gravitational fields - see $e.g.$ Refs.~\cite{Chowdhuri:2020ipb,Badia:2022phg,Gao:2023mjb,dePaula:2023ozi,Junior:2021dyw,Sengo:2022jif,Junior:2021svb,Novo:2024wyn,Xavier:2023exm,Hou:2022eev}. While weak lensing, such as the Sun’s deflection of light~\cite{Dyson:1920cwa}, provided the first observational test of GR, strong lensing occurs near ultra-compact objects, including BHs or horizonless alternatives. 

An ultra-compact object is defined as one where light can form circular trajectories, known as \textit{light rings} (LRs). A theorem~\cite{Cunha:2020azh} guarantees the existence of unstable LRs for BH spacetimes for each rotation sense; $i.e$ BHs are always ultra-compact. This result was later generalized for a configuration of $N$ collinear BHs, which must accommodate at least $N$ unstable LRs for each sense of rotation~\cite{Cunha:2024ajc}. Since this result relies solely on boundary conditions and not on the field equations, it applies to the TW solution as well. The topological techniques developed in Refs.~\cite{Cunha:2020azh,Cunha:2024ajc} offer a powerful framework for analyzing the LR structure of both BH spacetimes and horizonless objects~\cite{Cunha:2017qtt,Xavier:2024iwr}. 

The idea of bound orbits that neither fall into the BHs nor escape to infinity encompasses more than just the LRs. Although LRs are a prominent example, they belong to a broader class of null trajectories known as fundamental photon orbits (FPOs)~\cite{Cunha:2017eoe}. Within this set of
orbits, some (or all) are associated with points on the shadow edge of a BH. The BH shadow can serve as an important signature of its underlying spacetime, potentially enabling the identification of the specific type of BH being observed. However, there are cases where the shadows of totally different objects may exhibit similar patterns or even appear identical~\cite{Olivares:2018abq,Lima:2021las,Herdeiro:2021lwl,Rosa:2022tfv,Sengo:2024pwk}.

In this work, we explore the LRs, shadows, and gravitational lensing effects of the TW spacetime. As we will demonstrate, the LR structure—or more generally, the FPO structure—of this rotating, KK theory counterpart of the MP spacetime, is significantly more involved than that of its static counterpart. The remainder of this paper is organized as follows. In Sect.~\ref{Sec2}, we review key aspects of the single RL BH solution, which is the basis for deriving the TW solution, revisiting its main properties. Sect.~\ref{Sec3} reviews key aspects of the single TW solution restricted to the case of only two BHs in equilibrium. Sect.~\ref{Sec4} focuses on the motion of null geodesics in the TW spacetime, where we define the 2-dimensional effective potentials $\mathcal{H}_\pm$ and analyze the LR structure using the techniques introduced in Ref.~\cite{Cunha:2020azh,Cunha:2024ajc}. In Sect.~\ref{Sec5}, we present our findings on the shadow and gravitational lensing effects of the TW spacetime. Finally, Sec.~\ref{Sec6} presents  our final remarks.

\section{Single BH solution: Rasheed-Larsen spacetime}
\label{Sec2}
\subsection{KK theory}
In this section we review some aspects of the RL spacetime, a solution in KK theory. 

KK theory emerges from Einstein's pure gravity theory in higher-dimensional spacetimes. Specifically, we consider $(1+4)$-dimensional vacuum GR, described by the action:
\begin{equation}
\label{action5}S=\frac{1}{16\pi G_5}\int d^5X\sqrt{-{g}_{(5)}}{R_{(5)}} ,
\end{equation}
where $G_5$ is Newton's constant in five spacetime dimensions, $g_{(5)}$ and $R_{(5)}$ denote the five dimensional metric and Ricci scalar, respectively, and $X^M=(x^\mu,x^5)$ are the five dimensional coordinates.
The fifth dimension, $x^5$, is compactified, and the 5-dimensional metric remains invariant under translations along this compact dimension. These conditions allow the 5-dimensional vacuum equations to be mapped into a 4-dimensional theory with a (dilaton) scalar field coupled to Maxwell electrodynamics, under the KK ansatz
\begin{equation}
d{s}^2_{(5)}=e^{\phi/\sqrt{3}}g_{\mu\nu}dx^\mu dx^\nu + e^{-2\phi/\sqrt{3}}(dx^5+2A_\mu dx^\mu)^2  ,    
\label{kkansatz}
\end{equation}
leading to KK theory
\begin{align}
\label{KKtheory}
S=\frac{1}{16\pi G_4}\int d^4x\sqrt{-g}\left[R-\frac{1}{2}(\nabla{\phi})^2-\frac{1}{4}e^{-\sqrt{3}{\phi}}{F}^2 \right] ,
\end{align}
where $\phi$ is the dilaton field, $F^2= F_{\mu\nu}F^{\mu\nu}$,  $F_{\mu\nu}=\partial_\mu A_\nu-\partial_\nu A_\mu$ is the Maxwell 2-form and $A_\mu$ the potential 1-form.
Thus, the higher-dimensional geometry naturally incorporates both scalar and electromagnetic sources within a 4-dimensional framework. This corresponds to a specific type of Einstein-Maxwell-dilaton theory, where the dilaton coupling constant is fixed as $\sqrt{3}$, by the process of dimensional reduction.

\subsection{5-dimensional metric}
\begin{figure}[h!]
	\centering
	\includegraphics[width=\columnwidth]{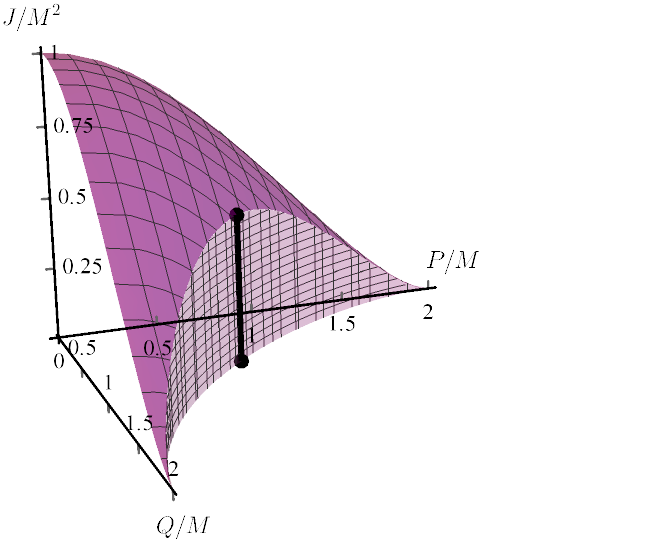}
	\caption{Surfaces of extreme solutions in KK theory. The under-rotating extremal solutions fall into  the lighter purple wall. Fixing the charges and varying $J$, one gets a vertical line segment, as illustrated by the black solid line, with a vanishing minimum value of $J$. The other class of extremal solutions fall into  the darker purple surface.}
	\label{Extremal}
\end{figure}

The RL BH represents a stationary and axisymmetric solution in KK theory~\eqref{KKtheory}, characterized by a connected event horizon. This solution is parametrized by four independent physical quantities: the mass $M$, angular momentum $J$, electric charge $Q$, and magnetic charge $P$. For such family, there exists two distinct \textit{extremal} classes, each corresponding to a different condition that saturates the bound for a horizon to exist. These two conditions can be represented within the parameter space shown in Fig.~\ref{Extremal}. 

The TW solution is constructed using one of these classes, called \textit{under-rotating extremal solutions}, for which $Q$ and $P$ are non-vanishing and $J$ can take values in a certain (charge) dependent interval, with the minimum value being always zero. This is illustrated in Fig.~\ref{Extremal}.

Using a spherical-like coordinate system $(t,r,\theta,\varphi,x^5)$, the under-rotating limit of RL extremal solution~\mbox{\cite{Rasheed:1995zv,Larsen:1999pp}} is described (in their five dimensional guise) by the 5-dimensional metric
\begin{equation}
	\begin{aligned}
		ds_{(5)}^2=&\frac{H_2}{H_1}\biggl\{dx^5-\left[2\left(r+\frac{p}{2}-pj\cos\theta\right)\right]\frac{Q}{H_2}dt  \\
		&-\left[2H_2\cos\theta-q\left(r+\frac{pq}{p+q}\right)j\sin^2\theta\right]\frac{P}{H_2}d\varphi\bigg\}^2\\
		&-\frac{r^2}{H_2}\left(dt+\frac{2jPQ\sin^2\theta d\varphi}{r}\right)^2\\
		&+H_1\left(\frac{dr^2}{r^2}+d\theta^2+\sin^2\theta d\varphi^2\right),
	\end{aligned}
\end{equation}
where
\begin{equation}
	H_1=r^2+r p+\frac{p^2 q (1+j\cos\theta)}{2(p+q)},
\end{equation}
\begin{equation}
	H_2=r^2+r q+\frac{p q^2 (1-j\cos\theta)}{2(p+q)},
\end{equation}
and $p$ and $q$ are positive quantities related to the electric and magnetic charge by
\begin{equation}
	P^2=\frac{p^3}{4(p+q)},\ \ \ \ Q^2=\frac{q^3}{4(p+q)}.
\end{equation}
The mass and angular momentum are~\cite{Teo:2023wfd} $4M=p+q$ and $J=jPQ$ and are constrained by $(P/M)^{2/3}+(Q/M)^{2/3}=2^{2/3}$ and $|J|<|PQ|$. 
Following~\cite{Teo:2023wfd}, to construct the multi-centered solution, a simplification is achieved by considering $P=Q=M/\sqrt{2}$, yielding the simplified five dimensional geometry
\begin{equation}\label{ds5e}
	\begin{aligned}
		ds^2_{(5)}&=\frac{H_2}{H_1}\bigl\{dx^5-\sqrt{2}H_2^{-1}[M(r+M)-2J\cos\theta]dt\\
		&-\sqrt{2}\left[M\cos\theta-2J H_2^{-1}(r+M)\sin^2\theta\right]d\varphi\bigr\}^2\\
		&-\frac{r^2}{H_2}\left(dt+\frac{2J\sin^2\theta d\varphi}{r}\right)^2\\
		&+H_1\left(\frac{dr^2}{r^2}+d\theta^2+\sin^2\theta d\varphi^2\right)
	\end{aligned},
\end{equation}
where now $H_{1,2}=(r+M)^2\pm2 J \cos\theta$. While imposing equality between the charges is not a strict requirement, it greatly simplifies the construction of the multi-centered solution. Eq.~\eqref{ds5e} describes a BH solution that lies along the black line segment on the parameter space represented in Fig.~\ref{Extremal}. This solution is characterized by two parameters, $M$ and $J$, where $|J| < J_{\text{e}} = M^2 / 2$. Here, $J = 0$ corresponds to the bottom black point in Fig.~{Extremal}, while $J = J_{\text{e}}$ represents the point on the top.

\subsection{4-dimensional metric}

After performing dimensional reduction with~\eqref{kkansatz}, the particular case of equal charges BHs, Eq.~\eqref{ds5e}, leads to the a four-dimensional spacetime, characterized by its metric, gauge potential, and scalar field configuration:
\begin{equation}\label{R1}
	\begin{aligned}
		ds^2=&-\frac{(r-M)^2}{\sqrt{r^4-4J^2\cos^2\theta}}\left(dt+\frac{2J\sin^2\theta d\varphi}{r-M}\right)^2\\
		&\sqrt{r^4-4J^2\cos^2\theta}\left[\frac{dr^2}{(r-M)^2}+d\theta^2+\sin^2\theta d\varphi^2\right],
	\end{aligned}
\end{equation}
\begin{equation}\label{R2}
	\begin{aligned}
		A_\mu dx^\mu=-\sqrt{2}\bigg[\bigg(\frac{M r-2 J\cos\theta}{r^2-2J\cos\theta}\bigg)&dt+\bigg(M\cos\theta\\&-\frac{2Jr\sin^2\theta}{r^2-2J\cos\theta}\bigg)d\varphi\bigg],
	\end{aligned}
\end{equation}
\begin{equation}\label{R3}
	\phi=\frac{\sqrt{3}}{2}\ln\left(\frac{r^2+2J\cos\theta}{r^2-2J\cos\theta}\right),
\end{equation}
respectively.

The event horizon is located at $r = M$, enclosing a curvature singularity at $r = \sqrt{2 |J \cos \theta|}$. In the limit $J = 0$ the dilaton trivializes, and this solution reduces to the extremal RN BH of Einstein-Maxwell theory.  

\begin{figure}[h!]
	\centering
	\includegraphics[width=\columnwidth]{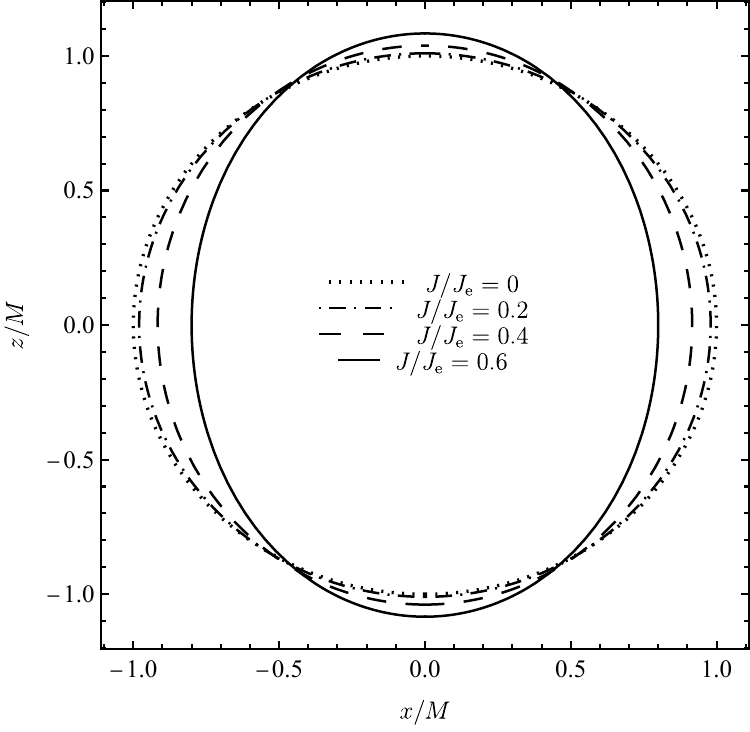}
	\caption{Isometric embedding of the horizon into 3-dimensional Euclidean space for $J/J_\text{e}=0,0.2,0.4,0.6$. As the angular momentum increases, the area diminishes, in accordance with the expression of $A_H$. In the limit of $J/J_\text{e}\to 1$, the area $A_H$ vanishes.}
	\label{HorizonTW}
\end{figure}

The geometry of the event horizon is determined by restricting Eq.~\eqref{R1} to the 2-surface  $t = \text{constant}$ and $r = M$. The area $A_H$ of this surface is given by $A_H=4\pi\sqrt{M^4-4J^2}$. The horizon Gaussian curvature is everywhere positive throughout the entire parameter space. Consequently, there exists a unique isometric embedding (up to rigid rotations) of the horizon surface into a 3-dimensional Euclidean space~\cite{berger2003panoramic}. In Fig.~\ref{HorizonTW} we display the isometric embedding for $J/J_\text{e}=0,0.2,0.4,0.6$.

The parameters $M$ and $J$ are constants obtained via Komar integrals evaluated at infinity. However, we may also evaluate the mass and angular momentum by Komar integrals over the horizon~\cite{Delgado:2016zxv}. From this, we find that the horizon mass $M_H$ and horizon angular momentum $J_H$ are given by
\begin{equation}
	M_H=0,
\end{equation}
\begin{equation}
	J_H=\frac{\left(4 J^2-M^4\right) \left[\left(4 J^2-M^4\right) \tanh ^{-1}\left(\frac{2 J}{M^2}\right)+2 J M^2\right]}{16 J^2 M^2},
\end{equation}
respectively.

\begin{figure}[h!]
	\centering
	\includegraphics[width=\columnwidth]{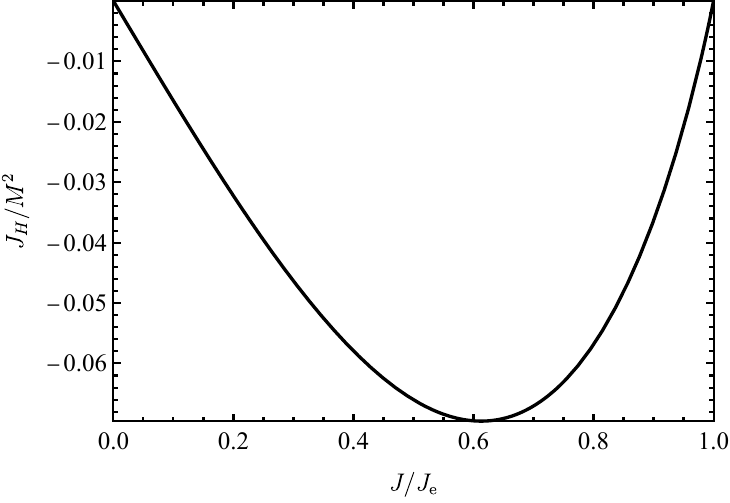}
	\caption{Angular momentum at the horizon $J_H$, as a function of the parameter $J$ of the solution~\eqref{R1}.}
	\label{J}
\end{figure}

Thus, RL under-rotating extremal solutions have their entire mass contained outside the horizon, carried by the external fields, a characteristic shared with the RN extremal solution~\cite{Delgado:2016zxv}, while $J_H\neq0$ in general. Figure~\ref{J} shows the profile of $J_H$ as a function of $J$, from which we observe that the horizon angular momentum has an opposite sign to the angular momentum of the whole spacetime.

Despite the non-vanishing horizon angular momentum, we emphasize that the horizon has zero angular velocity, as inferred from $-g_{t\varphi}/g_{\varphi\varphi}=2J(r-M)/(r^4-4J^2)$~\cite{Rasheed:1995zv,Larsen:1999pp}.  This implies the absence of an ergoregion. Unlike other well known cases, as the BMPV BH in $D=5$, in which the absence of an ergoregion can be associated to supersymmetry~\cite{Gibbons:1999uv}, this class of solutions is not supersymmetric~\cite{Teo:2023wfd}.




\section{Teo-Wan spacetime}
\label{Sec3}
\subsection{The solution}

It was remarked in Ref.~\cite{Teo:2023wfd} that the RL solution in the under-rotating extremal limit is contained within a class of solutions reported by Clément~\cite{Clement:1986bt},  characterized by two harmonic functions. This structure is the trademark of a superposition principle and allowed Teo and Wan to extend the single extremal BH solution~\eqref{ds5e} to a multi-BH configuration by generalizing the single center harmonic functions to multi-center ones. 
The resulting TW solution describes an asymptotically flat, stationary, axisymmetric, multi-centered nonsingular equilibrium configuration of rotating dyonic BHs, each with an extremal horizon. The particular case of two BHs may be written as
\begin{equation}\label{TW1}
	ds^2=-\frac{(dt+\omega_\varphi^0 d\varphi)^2}{\sqrt{H_+ H_-}}+\sqrt{H_+ H_-}\left(d\rho^2+dz^2+\rho^2 d\varphi^2\right),
\end{equation}
\begin{equation}\label{TW2}
	\begin{aligned}
		A_\mu dx^\mu=\frac{\sqrt{2}}{H_-}\Bigg\{-&\left[(1+f)f-2g\right]dt\\&+\Bigg[(1+f)\omega_\varphi^0+\frac{H_- \tilde{\omega}_\varphi^5}{\sqrt{2}}\Bigg]d\varphi\Bigg\},
	\end{aligned}
\end{equation}
\begin{equation}\label{TW3}
	\phi=\frac{\sqrt{3}}{2}\ln\frac{H_+}{H_-},
\end{equation}
where
\begin{equation}
	H_{\pm}=(1+f)^2\pm 2g,
\end{equation}
\begin{equation}
	\omega_\varphi^0=\frac{2J_1\rho^2}{[\rho^2+(z+a)^2]^{3/2}}+\frac{2J_2\rho^2}{[\rho^2+(z-a)^2]^{3/2}},
\end{equation}
\begin{equation}
	\tilde{\omega}_\varphi^5=-\frac{\sqrt{2}M_1(z+a)}{\sqrt{\rho^2+(z-a)^2}}-\frac{\sqrt{2}M_2(z-a)}{\sqrt{\rho^2+(z+a)^2}},
\end{equation}
 and 
\begin{equation}
	f=\frac{M_1}{\sqrt{\rho^2+(z+a)^2}}+\frac{M_2}{\sqrt{\rho^2+(z-a)^2}},
\end{equation}
\begin{equation}
	g=\frac{J_1(z+a)}{\sqrt{\rho^2+(z+a)^2}^3}+\frac{J_2(z-a)}{\sqrt{\rho^2+(z-a)^2}^3}.
\end{equation}

The double BH spacetime is determined by 5 parameters, namely: $M_1,\ M_2,\ J_1,\ J_2$ and $a$, where $|J_1|<J_{\text{e},1}=M_1^2/2$, $|J_2|<J_{\text{e},2}=M_2^2/2$ and $a\geq0$.
The multi-centered solution reduces to the single 4-dimensional BH solution of Sec.~\ref{Sec2} with the coordinate change $\{\rho=(r-M)\sin\theta,z=(r-M)\cos\theta\}$, when $a\to0$ and with the identifications $M_1\to M/2,\ M_2\to M/2,\ J_1\to J/2,$ and $J_2\to J/2$. 

We remark that Eq.~\eqref{TW1}, expressed in cylindrical coordinates $ \{t, \rho, z, \varphi\} $, has a spatial sector conformal to the Euclidean 3-space, $ \mathbb{E}^3 $. Accordingly, we define the Euclidean position $ \boldsymbol{x} = (x, y, z) $, where $ x $, $ y $, and $ z $ are rectangular coordinates. The two BHs are located at $ \boldsymbol{x}_1 = (0, 0, -a) $ and $ \boldsymbol{x}_2 = (0, 0, a) $, making $ a $ the coordinate distance from the origin of $ \mathbb{E}^3 $ to each gravitating center.

The functions $f$ and $g$ are solutions of the Poisson equation on $\mathbb{E}^3$, respectively, namely
\begin{equation}
	\Delta_{\mathbb{E}^3}f=4\pi M_1\delta(\boldsymbol{x}+a\hat{\boldsymbol{z}})+4\pi M_2\delta(\boldsymbol{x}-a\hat{\boldsymbol{z}}),
\end{equation}
\begin{equation}
	\Delta_{\mathbb{E}^3}g=4\pi \boldsymbol{J}_1\cdot\nabla_{\mathbb{E}^3}\delta(\boldsymbol{x}+a\hat{\boldsymbol{z}})+4\pi \boldsymbol{J}_2\cdot\nabla_{\mathbb{E}^3}\delta(\boldsymbol{x}-a\hat{\boldsymbol{z}}),
\end{equation}
where $\delta(\cdot)$ is the Dirac delta function, $\hat{\boldsymbol{z}}$ denotes the unit vector along the $z$-axis and the vectors $\boldsymbol{J}_1$ and $\boldsymbol{J}_2$ are defined as $\boldsymbol{J}_1 = J_1 \hat{\boldsymbol{z}}$ and $\boldsymbol{J}_2 = J_2 \hat{\boldsymbol{z}}$. The operators $\Delta_{\mathbb{E}^3}$ and $\nabla_{\mathbb{E}^3}$ refer to the Laplacian and gradient on $\mathbb{E}^3$, respectively. Hence, $f$ represents a pair of monopole solutions with monopole strengths $M_1$ and $M_2$, while $g$ corresponds to a pair of dipole solutions with dipole moments $J_1$ and $J_2$.

The spacetime total mass, angular momentum and charge, calculated using Komar integrals and the Gauss's law, are given by $M_1+M_2,\ J_1+J_2$ and $\sqrt{2}(M_1+M_2)$, respectively. Moreover, the parameters $(M_1, J_1)$ and $(M_2, J_2)$ can be interpreted as the mass and angular momentum of the BHs located at $\boldsymbol{x}_1$ and $\boldsymbol{x}_2$, respectively, since Eqs.~\eqref{TW1}, \eqref{TW2}, and \eqref{TW3} reduce to Eqs.~\eqref{R1}, \eqref{R2}, and \eqref{R3} near each center, with the identifications $(M_1,J_1)\to(M,J)$ or $(M_2,J_2)\to(M,J)$. Therefore, all the analyses presented in Sec.~\ref{Sec2} regarding horizon quantities and horizon embedding, remain valid for the multi-centered solution.

\subsection{Majumdar-Papapetrou limit}

For $J_1 = J_2 = 0$, the scalar field $\phi$ vanishes, as $H_+ \big|_{J_1=J_2=0} = H_- \big|_{J_1=J_2=0}$. In this limit, the metric and the vector potential reduce to:

\begin{equation}\label{MP1}
	ds^2= \frac{dt^2}{(1+f)^2}+(1+f)^2(d\rho^2+dz^2+\rho^2d\varphi^2),
\end{equation}
\begin{equation}\label{MP2}
	A_\mu dx^\mu=-\frac{\sqrt{2}  f}{1+f}dt+ \tilde{\omega}_\varphi^5 d\varphi.
\end{equation}
The metric and gauge potential given in Eqs.~\eqref{MP1} and~\eqref{MP2} correspond to the dyonic MP solution\footnote{Usually the MP is presented with $A_t=(1+f)^{-1}$, different from Eq.~\eqref{MP2}. However, $-f/(1+f)=1/(1+f)-1$ and the $-1$ term can be gauged away, hence the two potentials are equivalent.}~\cite{Mazharimousavi:2010ma}.

\section{Null orbits}
\label{Sec4}





\subsection{Majumdar-Papapetrou null orbits}
\label{MPLRs}

This subsection reviews null orbits in the MP spacetime, with a particular focus on null planar orbits, including LRs and other general FPOs. Reference~\cite{Cunha:2020azh} introduced a robust method for detecting the presence of LRs within a given spacetime. This approach relies on evaluating a topological charge (TC), which is derived from a circulation integral involving the normalized gradient of the following potential functions:

\begin{equation}
	\mathcal{H}_\pm=\frac{-g_{t\varphi}\pm\sqrt{g_{t\varphi}^2-g_{tt}g_{\varphi\varphi}}}{g_{\varphi\varphi}}.
\end{equation}

The normalized gradients of $\mathcal{H}_\pm$ are defined by

\begin{equation}
	\textbf{v}_{\pm}=\left(\frac{1}{\sqrt{g_{\rho\rho}}}\frac{\partial \mathcal{H}_{\pm}}{\partial \rho}, \frac{1}{\sqrt{g_{zz}}}\frac{\partial \mathcal{H}_{\pm}}{\partial z}\right)=\left(\text{v}_\rho^\pm,\text{v}_z^\pm\right).
\end{equation}

To determine the location of each LR, one can find the critical points of the potentials $\mathcal{H}_\pm$, or equivalently, identify the zeros of the vector field $\textbf{v}_{\pm}$. By evaluating the circulation of $\textbf{v}_{\pm}$ along a closed path in the $(\rho, z)$-plane that encloses exactly one of these zeros, a TC can be assigned to each LR. The TC is defined as $+1$ if the winding direction of $\textbf{v}_{\pm}$ matches its circulation around the path; otherwise, it is $-1$. 

TC=$-1$ corresponds to a saddle point of $\mathcal{H}_\pm$, TC=$+1$ indicates either a maximum or a minimum. Whether the LR represents a maximum or a minimum, depends on the behavior of the vector field $\textbf{v}_\pm$ in a neighborhood of the point: if $\textbf{v}_\pm$ is convergent, the point is a maximum, if it is divergent, the point is a minimum. 

The total TC, calculated over a path that encloses all the LRs of a given spacetime, can be generically determined under appropriate boundary conditions. In Ref.~\cite{Cunha:2020azh}, it was shown that in any single BH spacetime with properties of stationarity, axial symmetry, circularity, and asymptotic flatness, the total TC for each potential $\mathcal{H}_{\pm}$ is always $-1$. This result was later generalized in Ref.~\cite{Cunha:2024ajc} to account for a configuration of $N$ collinear BHs, where the corresponding total TC is given by $-N$.

The MP LR structure can be analyzed through the lens of the TC formalism. The potential functions $\mathcal{H}_\pm^{\text{MP}}$ for the MP spacetime are given by

\begin{equation}\label{HpmMP}
	\mathcal{H}_{\pm}^{\text{MP}}=\frac{\pm1}{\rho  \left(1+f\right)^2}.
\end{equation} 

Its corresponding vector fields $\textbf{v}_{\pm}^{\text{MP}}$ are given by
\begin{equation}
	\left(\textbf{v}_{\pm}^{\text{MP}}\right)_j=\pm\frac{(1+f)\delta^\rho_j+2 \rho  \partial_j f}{\rho ^2 (1+f)^4},
\end{equation}
where $j=\rho,z$ and $\left(\textbf{v}_{\pm}^{\text{MP}}\right)_j$ denotes the $j$-th component of $\textbf{v}_{\pm}^{\text{MP}}$.

\begin{figure}[h!]
	\centering
	\includegraphics[width=\columnwidth]{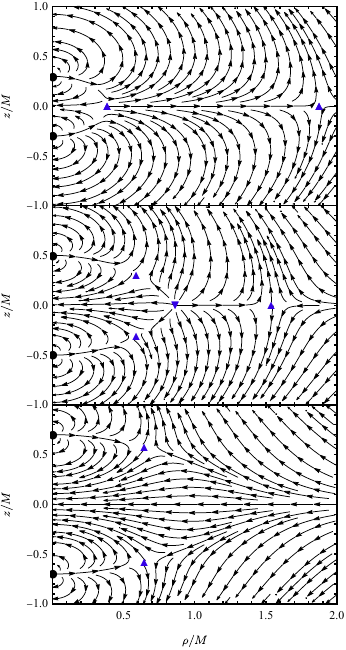}
	\caption{Vector field $\textbf{v}_{+}^{\text{MP}}$ over the $(\rho,z)$-plane for the 2-center MP with $a/M=0.3\text{ (top)},\ 0.5\text{ (middle)},\ 0.7 \text{ (bottom)}$. Blue upright triangles indicate the positions of LRs with a TC of $-1$, while the inverted triangle marks the location of the LR with a TC of $+1$.}
	\label{LRMP}
\end{figure}

Since the MP solution is static, the potentials $\mathcal{H}_{\pm}^{\text{MP}}$ and the vector fields $\textbf{v}_{\pm}^{\text{MP}}$ can only differ by a global minus sign. Hence it suffices to find the critical points and zeros of $\mathcal{H}_+^{\text{MP}}$ and $\textbf{v}_{+}^{\text{MP}}$, respectively. Fig.~\ref{LRMP} shows the LR configuration for $a/M =0.3,\ 0.5,\ 0.7$, along with the vector plot of $\textbf{v}_{+}^{\text{MP}}$ for equal-mass BHs.

In the MP spacetime, for $M_1=M_2=M$, equatorial LRs can only exist when the coordinate distance $a$ of each BH from the origin is below a critical value of $a_f/M \approx 0.5443$:
\begin{description}
    \item[i)] If $a$ is smaller than another critical distance, $a_i/M \approx 0.3849$, both LRs are saddle points~\cite{Shipley:2016omi}. 
    \item[ii)] Letting $a_i<a < a_f$, the internal LR becomes stable, while the external one remains a saddle point. Simultaneously, two off-equator LRs emerge, both saddle points.
     \item[iii)] For $a > a_f$, the off-equatorial LRs remain saddle points and the equatorial ones disappear. For a summary, see Table~\ref{LRtable}.
\end{description}

For $a_i<a < a_f$, the saddle point indicates the presence of two critical contour lines on the non-Killing submanifold, along which the LR exhibits stability (along one) and instability (along the other one). In general, these lines do not need to intersect orthogonally. However, it can be shown that for (stationary, axisymmetric, circular) $\mathbb{Z}_2$-symmetric spacetimes in four dimensions with an unstable LR at the equator, the critical lines are indeed orthogonal. This orthogonality arises from the $\mathbb{Z}_2$ symmetry, which enforces $\partial_z g_{\mu \nu} |{z=0} = \partial_z \partial_\rho g_{\mu \nu} |_{z=0} = 0\Rightarrow\partial_z \partial_\rho\mathcal{H}_+^{\text{MP}}|_{\text{LR}}=0$. Since the 2-center MP solution with equal masses is $\mathbb{Z}_2$, the stability of equatorial LRs is fully determined by their stability along the orthogonal $\rho$ and $z$ directions, $i.e.$ $\partial^2_\rho\mathcal{H}_+^{\text{MP}}|_{\text{LR}}$ and $\partial^2_z\mathcal{H}_+^{\text{MP}}|_{\text{LR}}$.

\begin{table*}[t]
	\centering
	\begin{tabular}{|l|c|c|c|c|} 
		\cline{2-5}
		\multicolumn{1}{l|}{} & $a<a_i$   & $a_i<a<a_f$  & $a_f<a<\overline{a}_f$   & $\overline{a}_f<a$    \\ 
		\hline
		Equatorial internal              & $\rho\smile |\ z \frown$ \ (-1) & $\rho\smile|\ z \smile$ \ (+1) & – & –  \\ 
		\hline
		Equatorial external              & $\rho\frown|\ z\smile$ \ (-1) & $\rho\frown |\ z\smile$ \ (-1) & – & –  \\ 
		\hline
		Top (off-equator)                   & – & $\rho\smile |\ z\smile$ \ (-1)& $\rho\smile |\ z\smile$ \ (-1) &  $\rho\frown|\ z\smile$ \ (-1) \\ 
		\hline
		Bottom (off-equator)                & – & $\rho\smile |\ z\smile$ \ (-1) & $\rho\smile |\ z\smile$ \ (-1) &  $\rho\frown|\ z\smile$ \ (-1) \\ 
		\hline
	\end{tabular}
	\caption{LR (in)stability along the $\rho$ and $z$ directions of the MP solution with equal masses, for different ranges of the parameter $a$. Here, $\rho\smile$ and $z\smile$ denote stability, while $\rho\frown$ and $z\frown$ denote instability, in the respective directions. In brackets, after each LR, we present its TC. Clearly, in all cases the total TC is $-2$. But the analysis of purely the $\rho$ and $z$ behaviour may be misleading when cross derivatives of the potential are non-vanishing.}
	\label{LRtable}
\end{table*}

Off-equatorial LRs in the range $a_i < a < a_f$ are stable in both $\rho$ and $z$ directions; however in this case stability along both directions is not sufficient to conclude that these LRs correspond to minima of $\mathcal{H}_+^{\text{MP}}$. Specifically, for off-equatorial LRs, the cross derivative $\partial_z \partial_\rho \mathcal{H}_+^{\text{MP}} |_{\text{LR}}$ does not vanish, so that the stability must be determined by the Hessian. Since $\det(\partial_i \partial_j \mathcal{H}_+^{\text{MP}}) < 0$, these LRs are indeed saddle points. 

\begin{figure}[h!]
	\centering
	\includegraphics[width=\columnwidth]{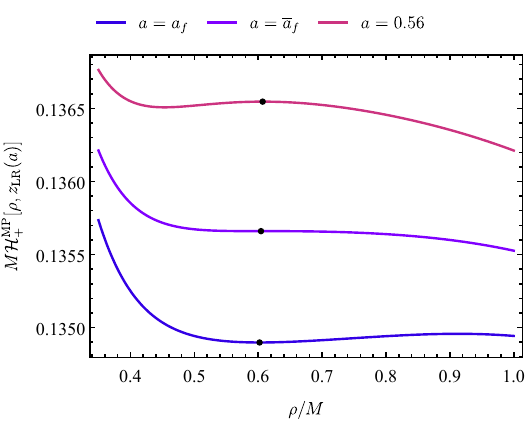}
	\caption{Plot showing the potential $\mathcal{H}{+}^\text{MP}$, as a function of the $\rho$ coordinate for $z = z_{\text{LR}}(a)$. The critical points associated with LRs are highlighted with black points.}
	\label{MPstabOffEqLR}
\end{figure}

It is worth noting that there is another critical value, $\overline{a}_f \approx 0.5515 M\gtrapprox a_f$, for the parameter $a$, at which the stability of the off-equatorial LRs along the $\rho$ direction transitions from stable to unstable. To better illustrate this stability transition, we plot the potential $\mathcal{H}_{+}^\text{MP}$ as a function of $\rho$ in Fig.~\ref{MPstabOffEqLR}, with $z = z_{\text{LR}}(a)$. Here, $z_{\text{LR}}(a)$ represents the $z$ coordinate of the top off-equatorial LR corresponding to the parameter $a$.

According to Table~\ref{LRtable}, the MP solution exhibits four LRs for any value of $a$ within the range $(a_i, a_f)$. Outside this range, there are only two LRs: equatorial when $a < a_i$, and off-equatorial when $a > a_f$. Since the TC is additive, the total TC is just the sum of the individual TCs. Hence, for all values of $a$, the MP binary has TC=$-2$, in accordance with Ref.~\cite{Cunha:2024ajc}.

\begin{figure}[h!]
	\centering
	\includegraphics[width=\columnwidth]{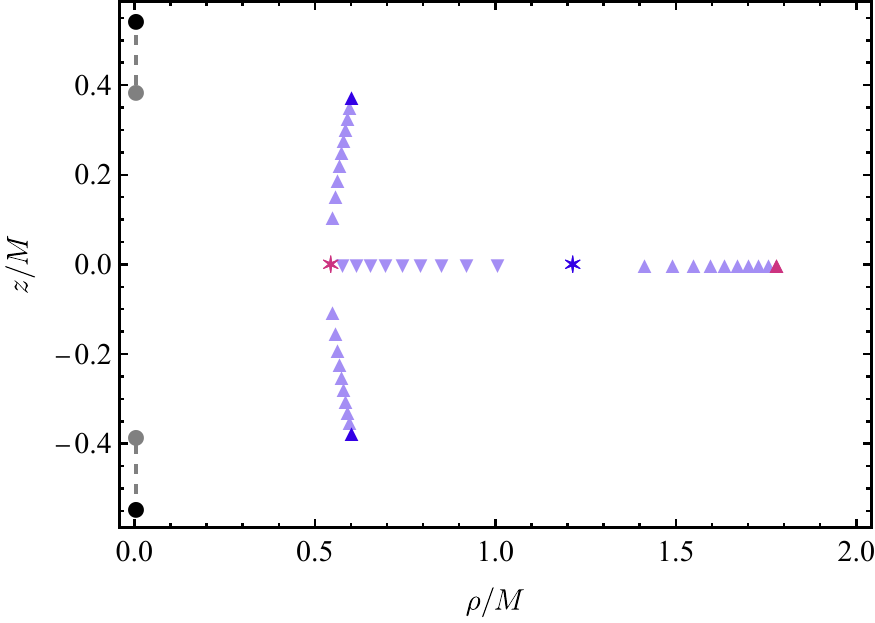}
	\caption{Positions of the LRs in the 2-center MP solution as the parameter $a$ varies. For $a = a_i$, the BHs are located at the gray circles. The corresponding LRs are represented by a pink six-pointed star and a triangle. The triangle denotes the external equatorial LR with  $\text{TC}=-1$, while the six-pointed star indicates a three-fold superposition of LRs, also with $\text{TC}=-1$. As the distance between the BHs increases, the six-pointed star splits into three distinct LRs: two are represented by purple upright triangles moving away from the equator, each with  $\text{TC}=-1$, and the third by purple inverted triangles moving along the equator with  $\text{TC}=+1$. For $a = a_f$, the BHs positions are represented by the black circle, the off-equatorial LRs reach the positions indicated by the dark blue triangles, and the two equatorial LRs merge at the position of the blue six-pointed star.}
	\label{LRcoal}
\end{figure}

\begin{figure}[h!]
	\centering
	\includegraphics[width=\columnwidth]{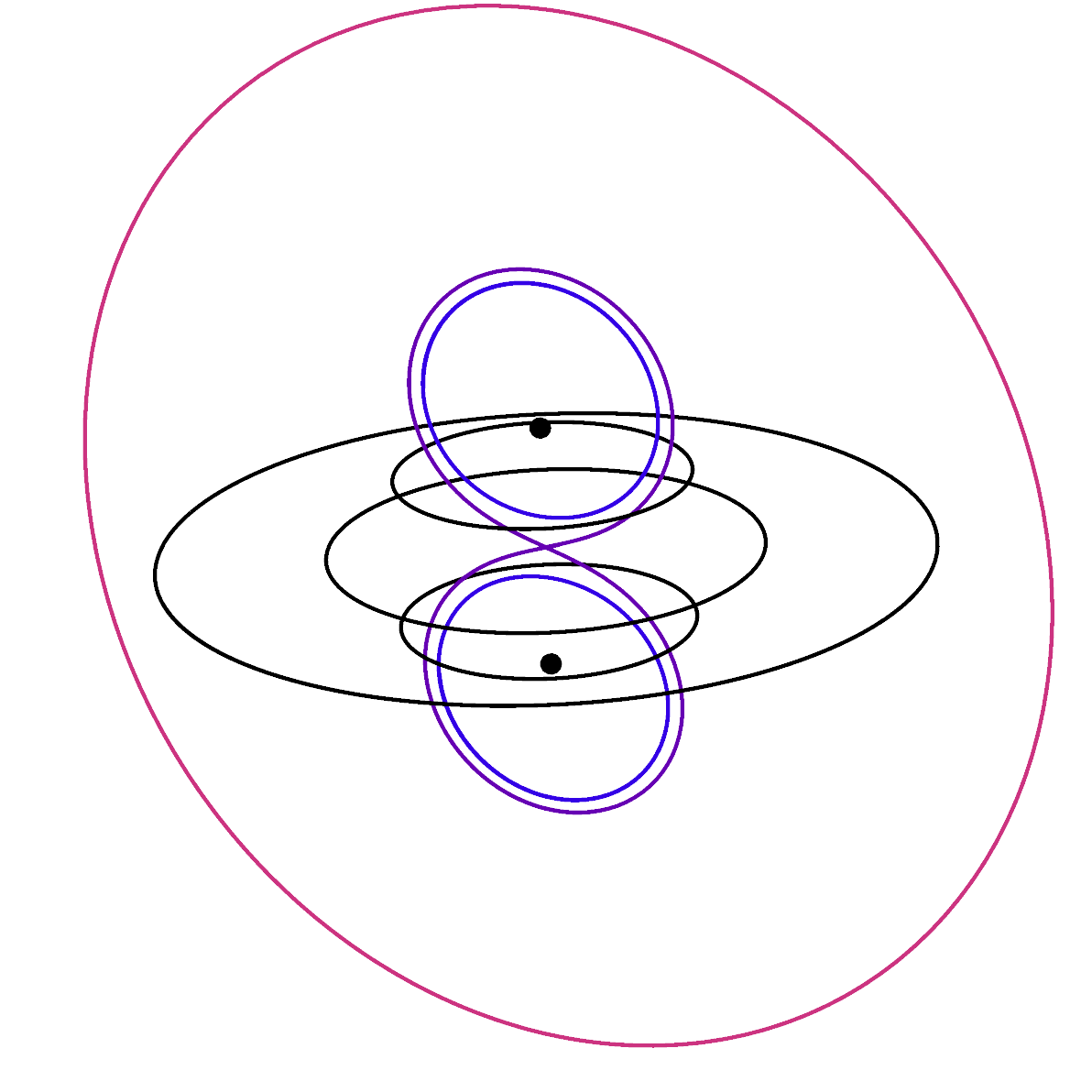}
	\caption{FPOs in the 2-center MP spacetime for $a/M = 0.5$. Black points indicate the BHs, black circles mark the LRs, as displayed in Fig.~\ref{LRMP} and colored lines show additional bounded null trajectories.}
	\label{MPFPOs}
\end{figure}

The variation in the number of LRs within the MP solution is associated with the coalescence of the $+1$ LR with one or more of the $-1$ LRs. As $a$ approaches $a_f$, the two equatorial LRs converge and annihilate each other due to their opposite TCs, resulting in the disappearance of LRs at the equator. Conversely, as $a$ approaches $a_i$, the $+1$ LR merges with the two off-equatorial plane LRs. The combined TC of these three merging LRs results in a single LR with a TC of $-1$. The interpolation between these two LR coalescing configurations can be visualized in Fig.~\ref{LRcoal}. In all the three cases, namely $a<a_i$, $a_i<a<a_f$ and $a_f$, the total TC is always equal to $-2$, 
which could not have been different, since the boundary conditions of the solution are not changed.

The LR positions in the closest configuration with $a < a_i$ closely resemble those found in a single BH solution, with the LRs confined to the equatorial plane. In contrast, when the BHs are taken apart beyond the critical distance $a_f$, the system exhibits a LR profile characteristic of two BHs, each with its own LR. The regime where $a_i < a < a_f$ represents a transitional phase between these two qualitatively distinct LR structures.

LRs represent a specific example of null trajectories within a larger family of orbits, termed FPOs. Broadly speaking, a FPO is a bound photon trajectory that neither falls into the BHs nor escapes to infinity~\cite{Cunha:2017eoe}. Such orbits for the MP spacetime were analyzed in Ref.~\cite{Shipley:2016omi}. In Fig.~\ref{MPFPOs}, we represent photon trajectories as curves in $\mathbb{E}^3$ using cylindrical coordinates $(\rho, z, \varphi)$. The LRs (black circles) are shown along with other null geodesics (colored lines) that lie within a vertical plane, i.e., $\varphi = 0$. The LRs depicted here correspond exactly to those in the second panel of Fig.~\ref{LRMP}.

\subsection{Teo-Wan null orbits}
\label{subsecIVB}

The LR structure of the TW solution is significantly more complex than that of the MP case discussed above. With the introduction of angular momentum into the system, the potentials $\mathcal{H}_\pm^{\text{TW}}$ for the TW spacetime no longer differ merely by a global sign, and are now given by

\begin{equation}
	\mathcal{H}_\pm^{\text{TW}}=\frac{1}{\pm\rho  \sqrt{H_+ H_{-}}-\omega_\varphi^0},
\end{equation}
and the associated vector field $\textbf{v}_{\pm}^{\text{TW}}$ can be written as
\begin{equation}
	\left(\textbf{v}_{\pm}^{\text{TW}}\right)_j=\frac{\sqrt{H_{+}H_{-}}\partial_j\omega_\varphi^0\mp\left[\delta^\rho_j+(\rho/2)\partial_j\right]H_+H_{-}}{ (H_{+} H_{-})^{3/4} \left(\omega_\varphi^0\mp\rho  \sqrt{H_{+} H_{-}}\right)^2},
\end{equation}
where $j=\rho,z$ and $\left(\textbf{v}_{\pm}^{\text{TW}}\right)_j$ denotes the $j$-th component of $\textbf{v}_{\pm}^{\text{TW}}$.

\begin{figure}[h!]
	\centering
	\includegraphics[width=\columnwidth]{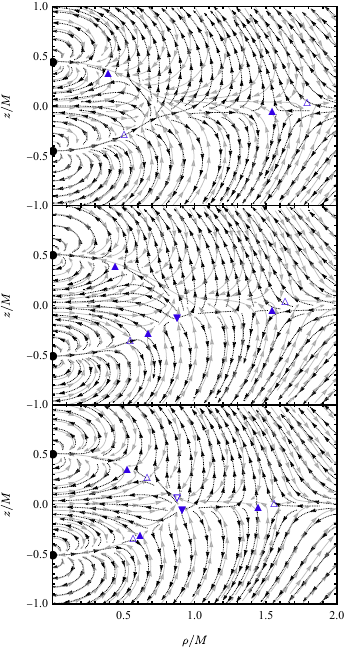}
	\caption{Vector fields $\textbf{v}_{+}^{\text{TW}}$ (dotted black) and $\textbf{v}_{-}^{\text{TW}}$ (gray) are shown on the $(\rho, z)$-plane with parameters $J_1 = 0$, $J_2 = J$, and $(a/M, J/M^2)$ values of $(0.45, 0.2)$ (top), $(0.51, 0.17)$ (middle), and $(0.51, 0.07)$ (bottom). Blue upright triangles mark the positions of LRs with a TC of $-1$, with filled triangles representing LRs obtained from $\mathcal{H}_+^{\text{TW}}$ and empty triangles from $\mathcal{H}_-^{\text{TW}}$. Similarly, the inverted triangles indicate the location of a LR with a TC of $+1$, with filled and empty markers denoting LRs from $\mathcal{H}_+^{\text{TW}}$ and $\mathcal{H}_-^{\text{TW}}$, respectively.}
	\label{LR_TW}
\end{figure} 

We remark that the total TC, for each $\mathcal{H}_\pm^\text{TW}$, remains the same as in the MP spacetime, regardless of the values of $J_1$ and $J_2$, as the boundary conditions are unchanged. Hence, although the LRs change position with different choices of angular momentum, their total TCs must still sum to $-2$.

In the MP case there is a maximum of 4 LRs. However, since the potentials $\mathcal{H}_\pm^\text{TW}$ are no longer related by a simple global sign, the TW spacetime may exhibit more than four LRs. In Fig.~\ref{LR_TW}, we show cases with 4, 6, and 8 LRs, along with the vector fields plot of $\textbf{v}_{\pm}^{\text{TW}}$ for equal-mass BHs, where one BH is non-rotating, while the other rotates.

\begin{figure}[h!]
	\centering
	\includegraphics[width=\columnwidth]{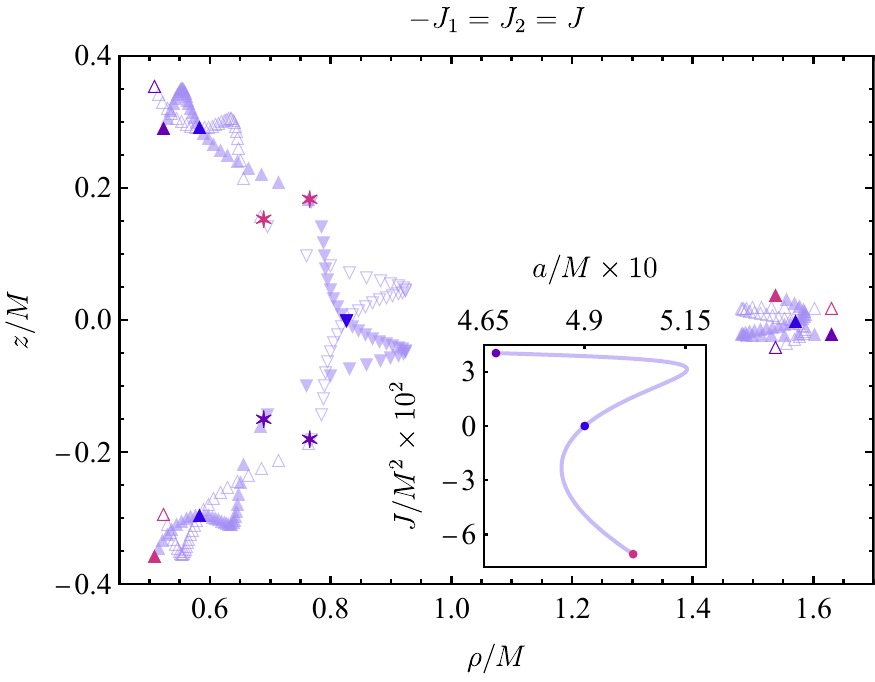}
	\caption{Variation in the positions of LRs of equal-mass, oppositely rotating BHs, with $-J_1 = J_2 = J$, as the parameters $a$ and $J$ change. Positions of LRs with a TC of $-1$ are indicated by upright triangles, where filled triangles correspond to those derived from $\mathcal{H}_+^{\text{TW}}$ and empty triangles to those from $\mathcal{H}_-^{\text{TW}}$. In contrast, inverted triangles mark the locations of LRs with a TC of $+1$, with filled and empty inverted triangles representing LRs from $\mathcal{H}_+^{\text{TW}}$ and $\mathcal{H}_-^{\text{TW}}$, respectively. The six-pointed stars represent superposition of LRs with opposite TCs.}
	\label{LR_TW_Traj}
\end{figure}

\begin{figure}[h!]
	\centering
	\includegraphics[width=\columnwidth]{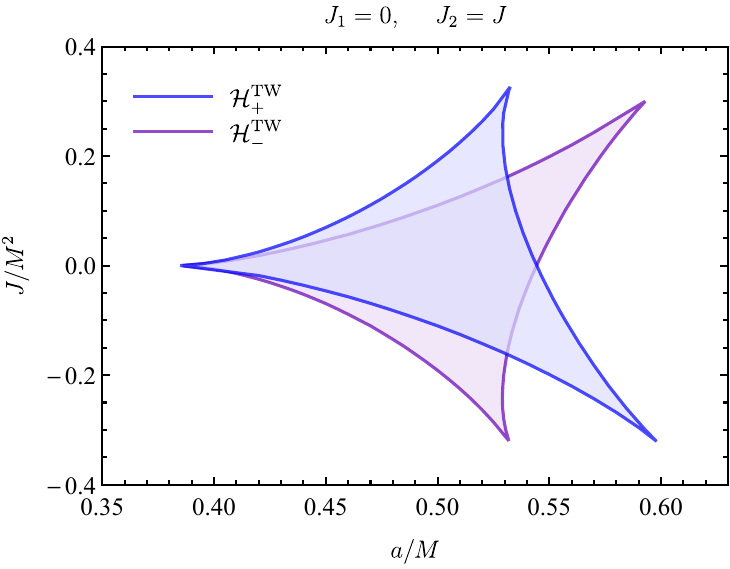}
	\includegraphics[width=\columnwidth]{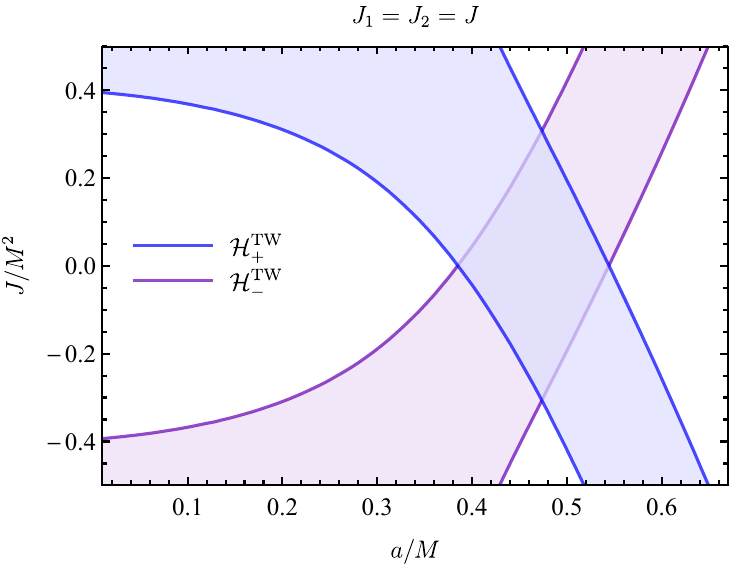}
	\includegraphics[width=\columnwidth]{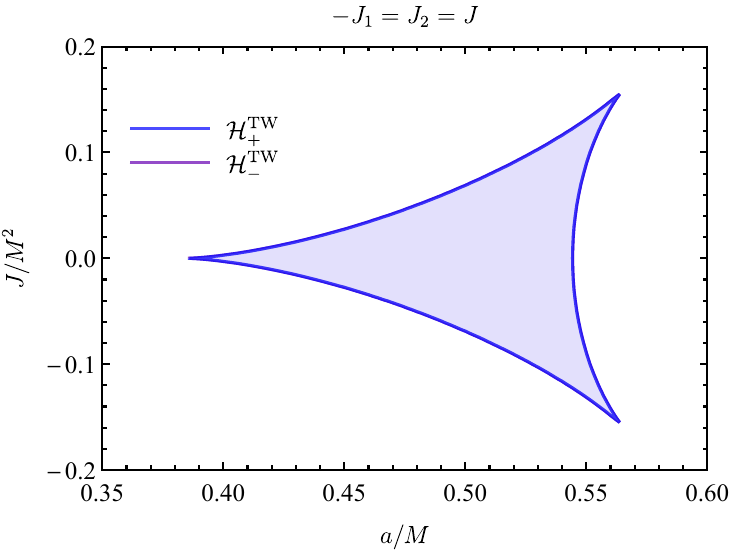}
	\caption{Parameter space highlighting regions with 4, 6 and 8 LRs for equal-mass BHs with $J_1=0,\ J_2=J$ (top); $J_1=J_2=J$ (middle) and $-J_1= J_2=J$ (bottom). In the white regions, the total number of LRs is 4. In the blue and purple regions without overlap, there are 6 LRs. In the overlapping blue and purple regions, the total number of LRs is 8.
	}
	\label{LR_Jcrit}
\end{figure}

In Fig.~\ref{LR_TW_Traj}, we present an analogous plot to Fig.~\ref{LRcoal} for the TW spacetime, showing how the positions of LRs change as we vary the solution parameters. We consider equal-mass, oppositely rotating BHs, $i.e.$ $M_1 = M_2 = M$ and $-J_1 = J_2 = J$. A curve in the parameter space $a \times J$ is shown in the inset of Fig.~\ref{LR_TW_Traj}. The corresponding LR locations are plotted on the $\rho \times z$ plane, with colors that match those in the parameter space. For example, the purple point in the parameter space corresponds to the LR configuration marked by the purple triangles and inverted triangles.

The TW solution allows for a minimum of 2 LRs and a maximum of 4 LRs for each potential, $\mathcal{H}_{\pm}^{\text{TW}}$. Consequently, the total number of LRs, for $J_1\neq0$ or $J_2\neq0$, will always be 4, 6, or 8. As in the MP case, changes in the number of LRs arise from the coalescence of LRs with different TCs. Given that the TW solution has 5 parameters, verifying all possible cases that result in 4, 6, or 8 LRs is significantly more challenging. Therefore, we restrict our analysis to equal-mass BHs and to three different configurations for angular momentum: $(i)J_1=0,\ J_2=J$; $(ii) J_1=J_2=J$ and $(iii)-J_1= J_2=J$. This approach leaves us with only two free parameters, namely $a$ and $J$.

\begin{figure*}[t]
	\centering
	\includegraphics[width=\columnwidth]{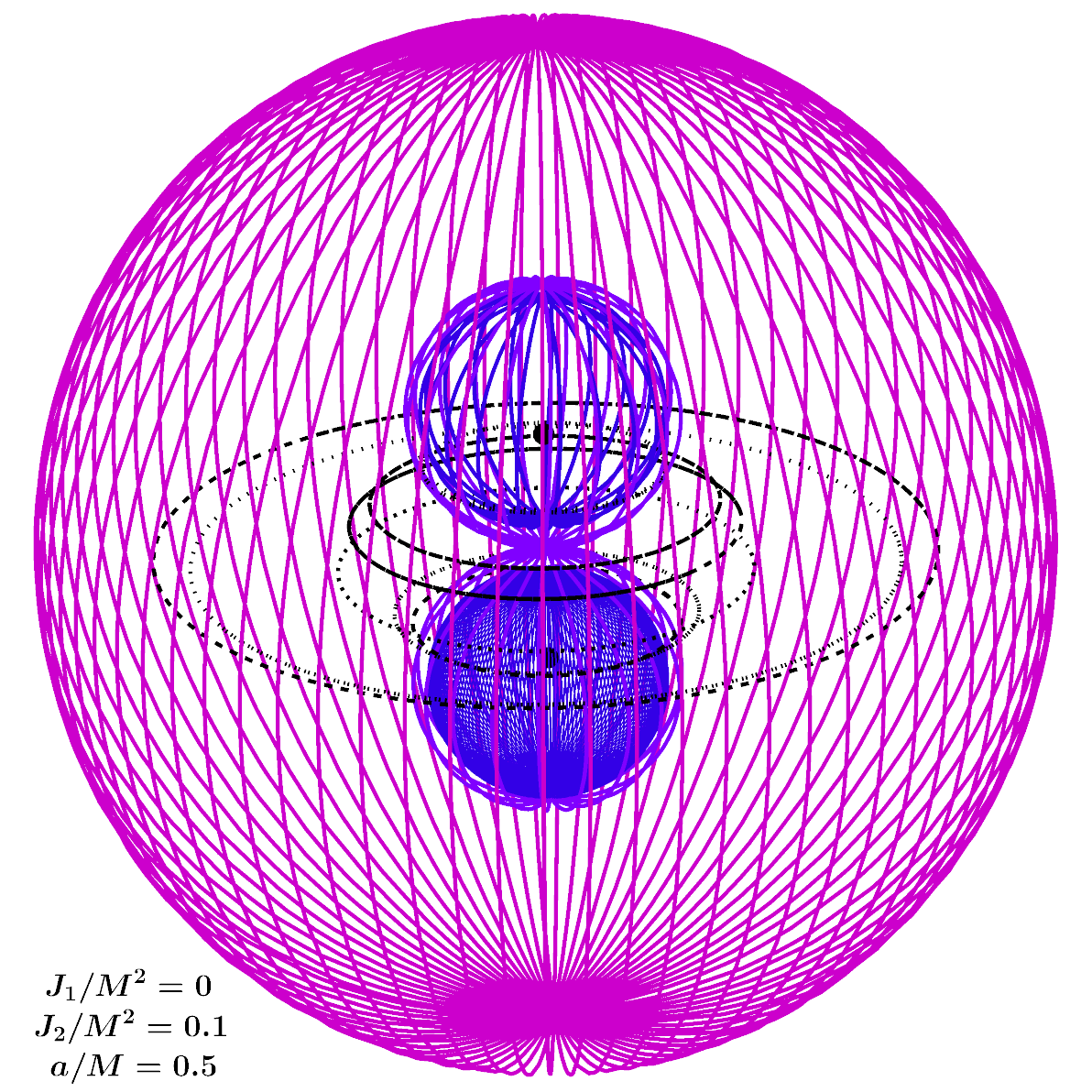}
	\includegraphics[width=\columnwidth]{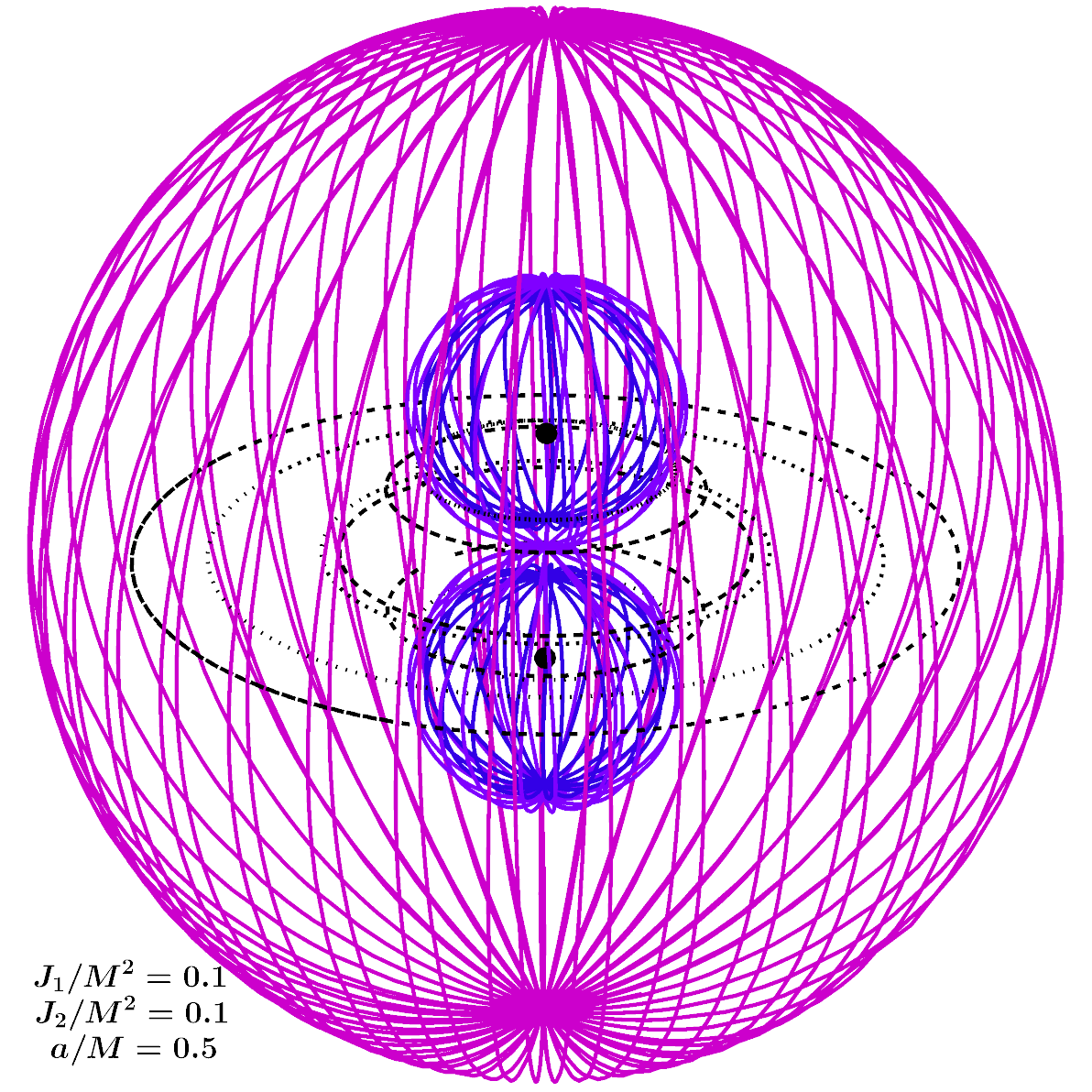}
	\includegraphics[width=\columnwidth]{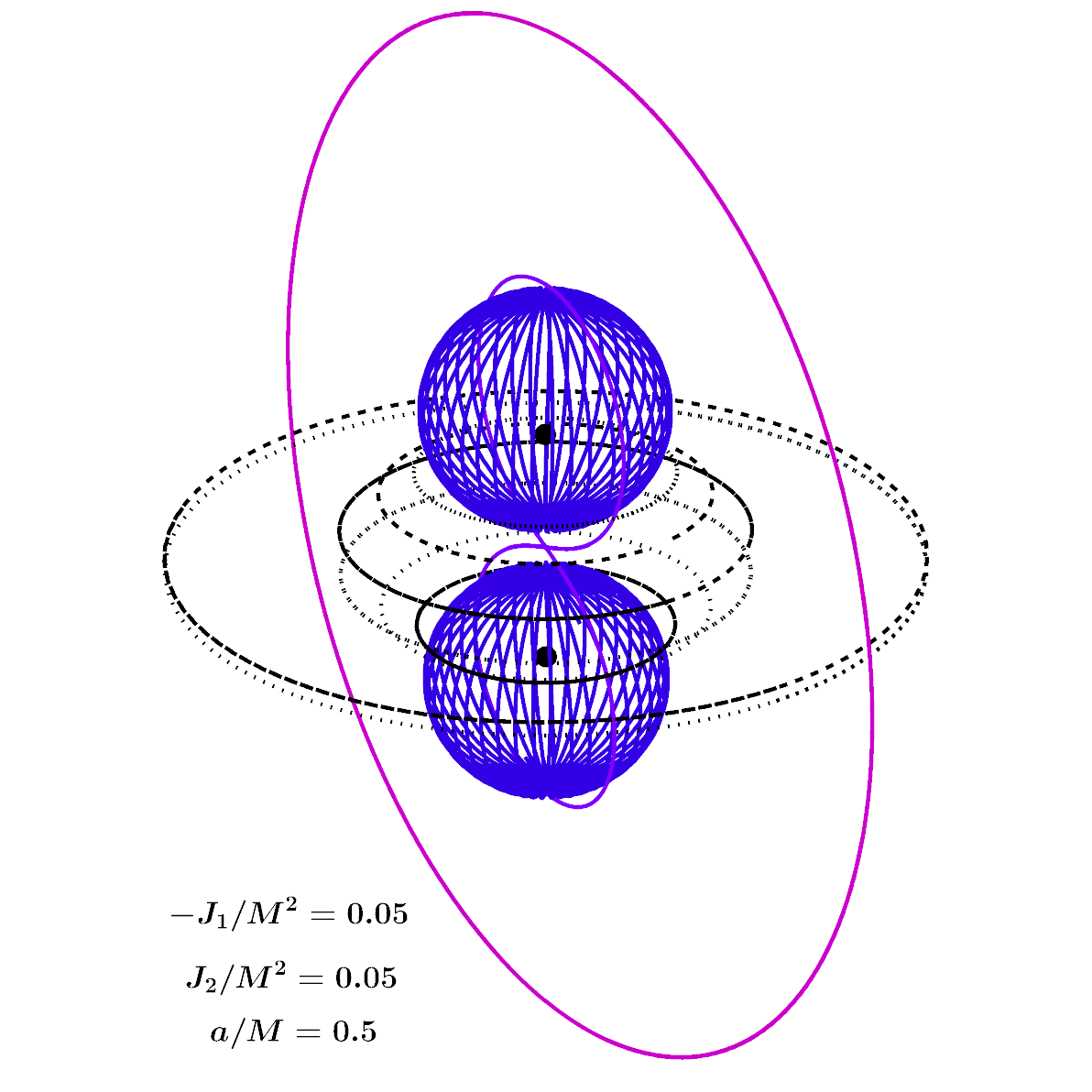}
	\caption{FPOs in the TW spacetime for $a/M = 0.5$ and three different angular momentum configurations. Black points indicate the BHs, black dashed (${H}_+^\text{TW}$) and dotted (${H}_-^\text{TW}$) orbits mark the LRs and colored lines, analogous to the colored lines exhibited in Fig.~\ref{MPFPOs}, show additional bounded null trajectories.
	}
	\label{TW_FPOs}
\end{figure*}

In Fig.~\ref{LR_Jcrit}, we show the regions of the parameter space corresponding to 4, 6, and 8 LRs (except for the trivial cases of $J=0$ and $a<a_i$, or $J=0$ and $a>a_f$, where we recover the 2 or 4 LRs of the MP solution discussed in Sec.~\ref{MPLRs}). The white regions indicate areas where both potentials, $\mathcal{H}_\pm^{\text{TW}}$, each contribute with 2 LRs. In the dark blue (light purple) region, $\mathcal{H}_+^{\text{TW}}$ ($\mathcal{H}_-^{\text{TW}}$) yields 4 LRs, while $\mathcal{H}_-^{\text{TW}}$ ($\mathcal{H}_+^{\text{TW}}$) produces only 2. In the overlapping regions, both potentials yield 4 LRs each.

For the case $J_1=0,\ J_2=J$, and $-J_1=J_2=J$ we notice that the boundary is composed of three different smooth pieces. This boundary represents critical values of $(a, J)$ where the number of LRs contributed by one potential changes from 2 to 4, or vice versa. Each smooth boundary segment corresponds to the coalescence of the $+1$ LR with one of the three $-1$ LRs.

For oppositely rotating BHs, the blue and purple regions coincide, even if the potentials $\mathcal{H}_\pm^{\text{TW}}$ are not identical. This overlap arises due to the symmetry of the parameter space along the horizontal axis $J=0$, which prevents configurations with a total of 6 LRs.

\begin{figure}[h!]
	\centering
	\includegraphics[scale=0.22]{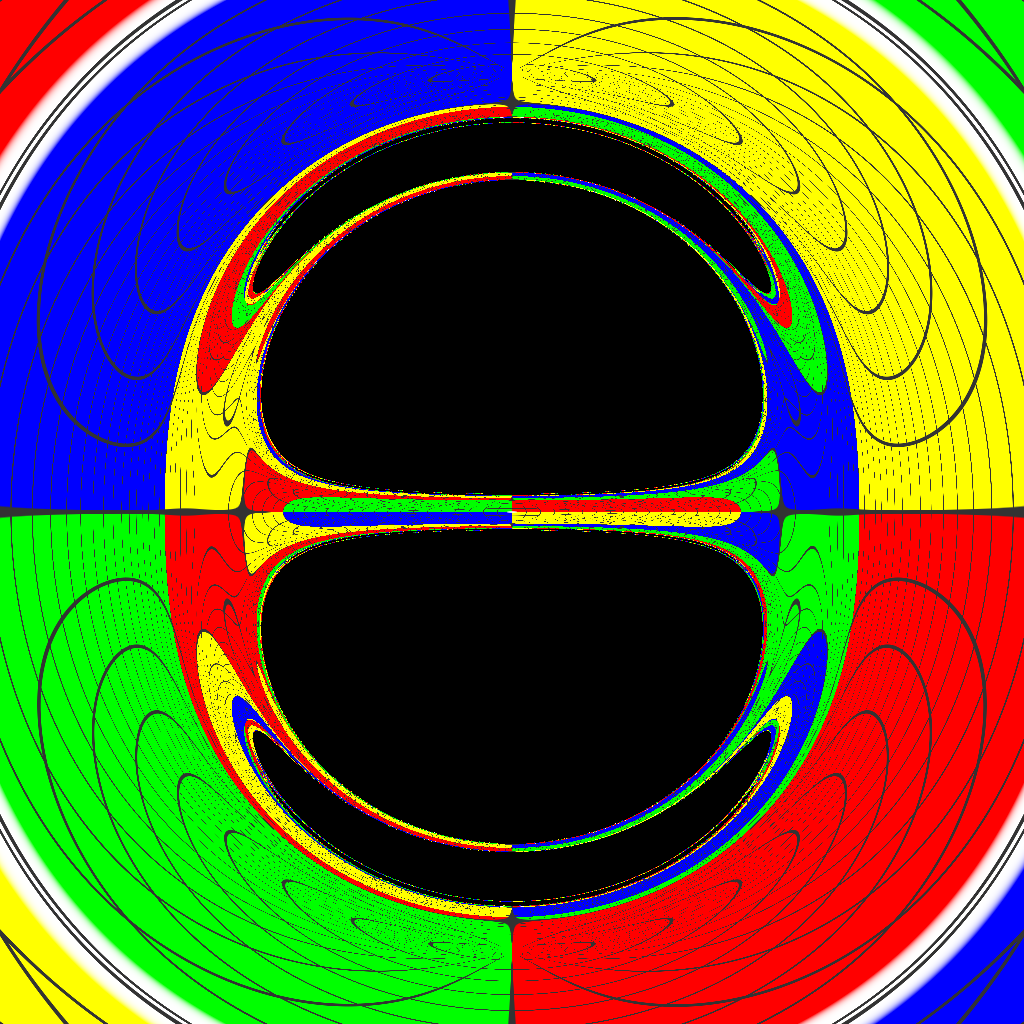}
	\caption{Shadow and gravitational lensing of the equal-mass MP solution. The observer is
		positioned at the equatorial plane ($z=0$) and at the radius $\rho = 15M$.}
	\label{MP_Shadow}
\end{figure}

\begin{figure*}[t]
	\centering
	\includegraphics[scale=0.22]{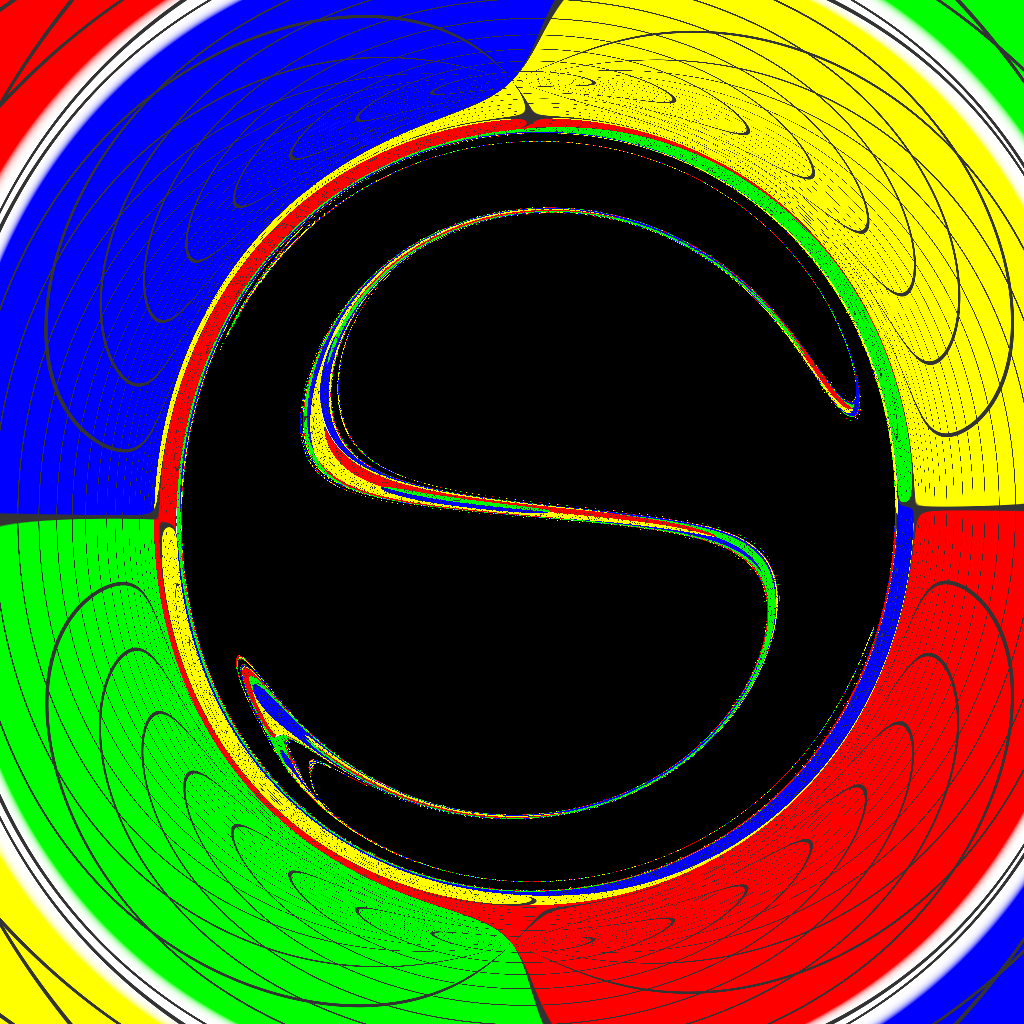}
	\includegraphics[scale=0.22]{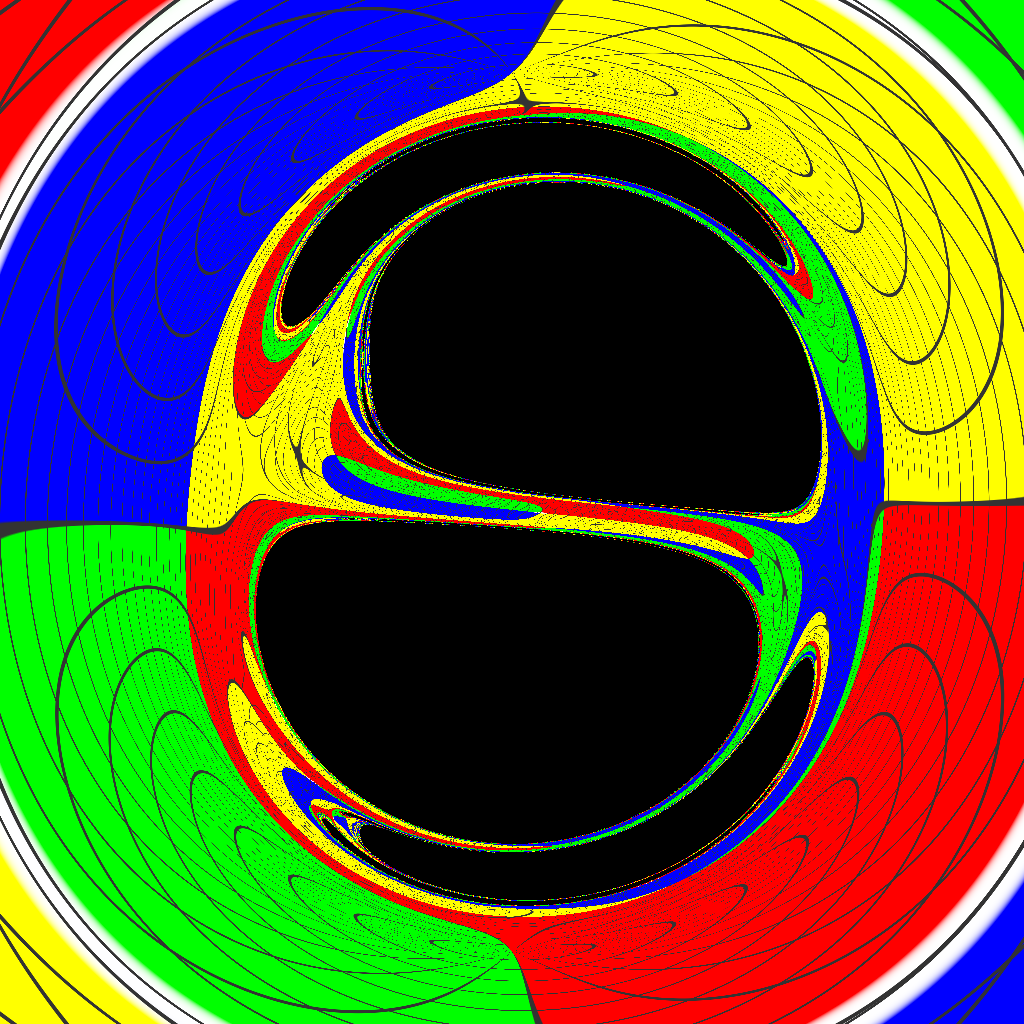}
	\includegraphics[scale=0.22]{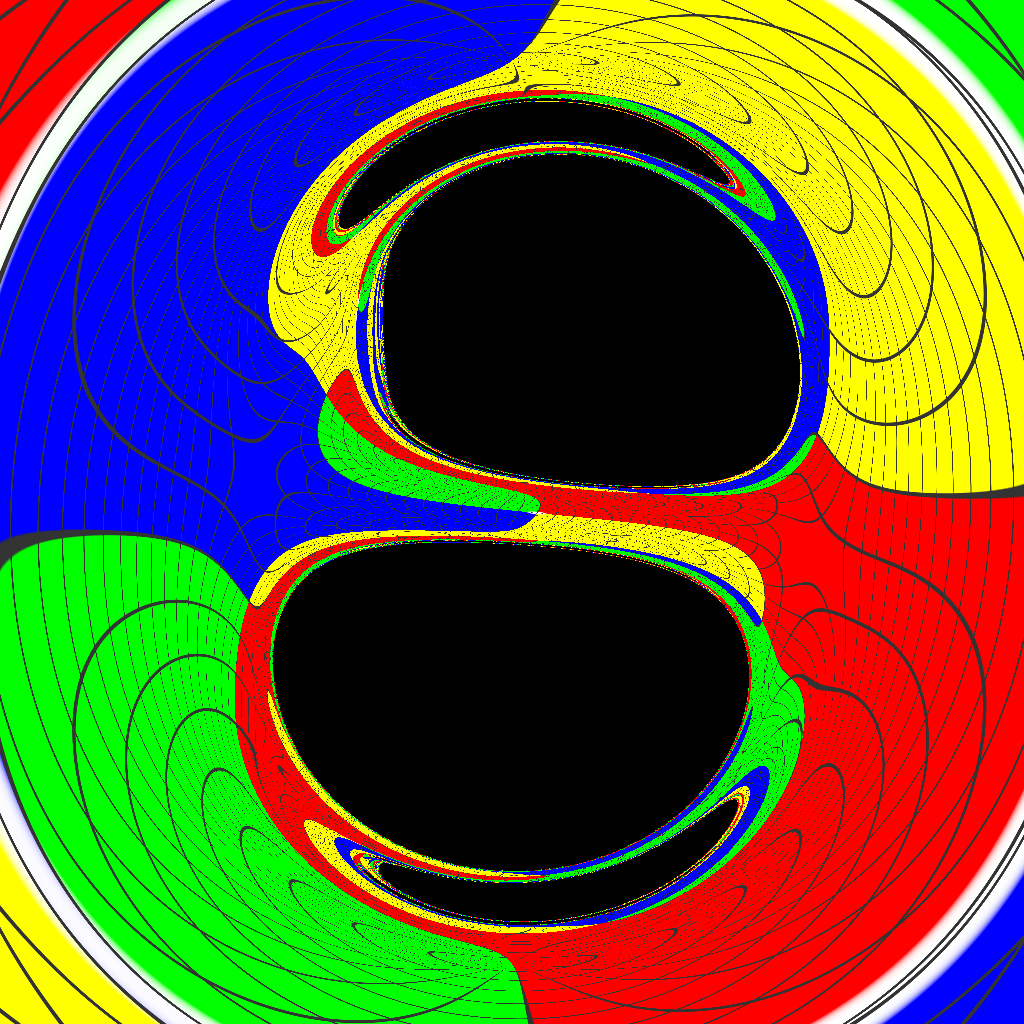}
	\includegraphics[scale=0.22]{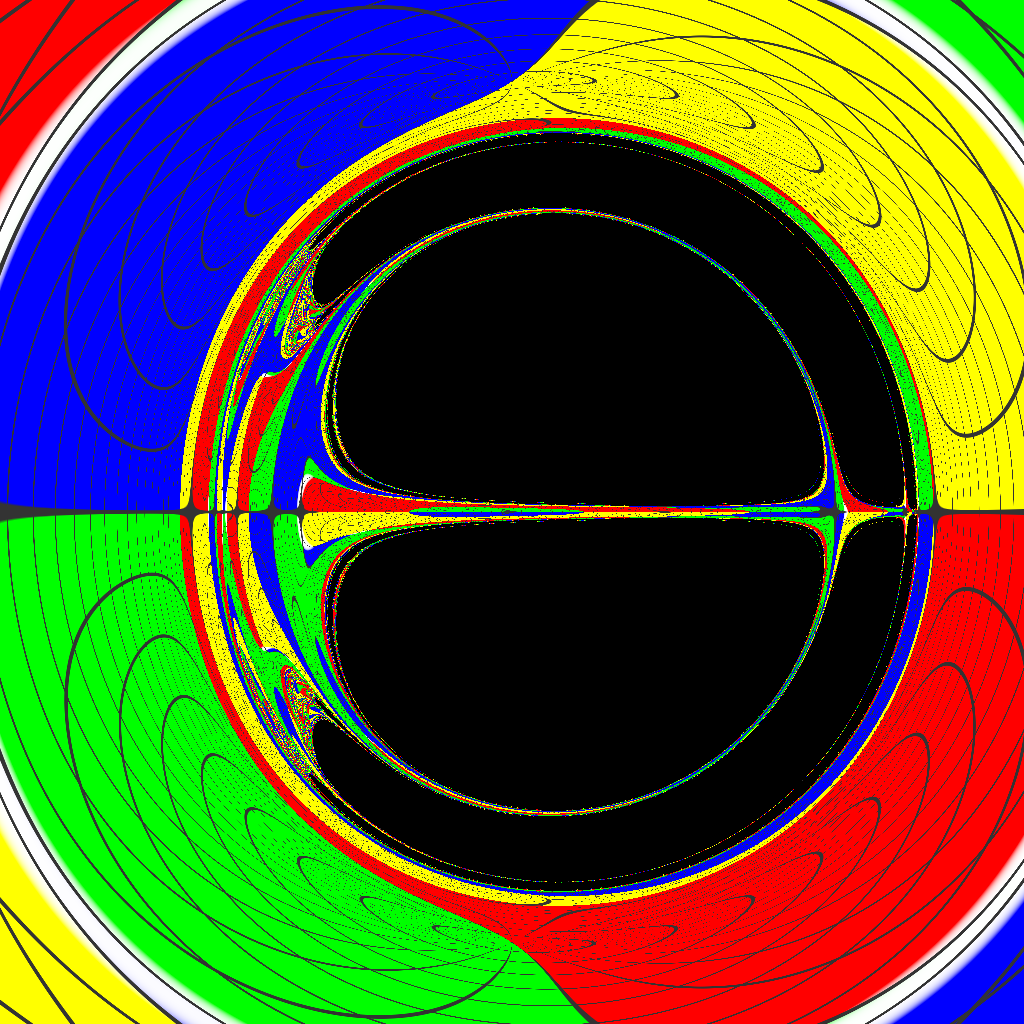}
	\includegraphics[scale=0.22]{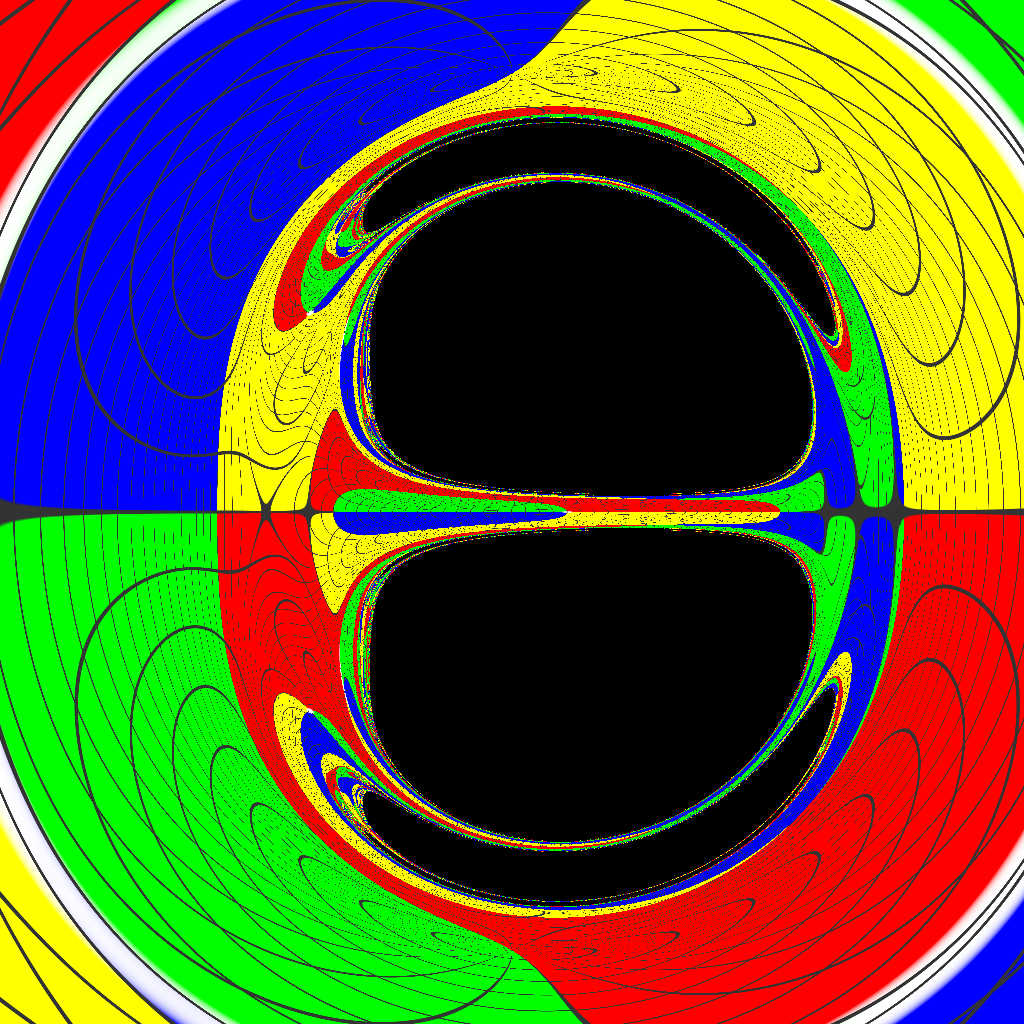}
	\includegraphics[scale=0.22]{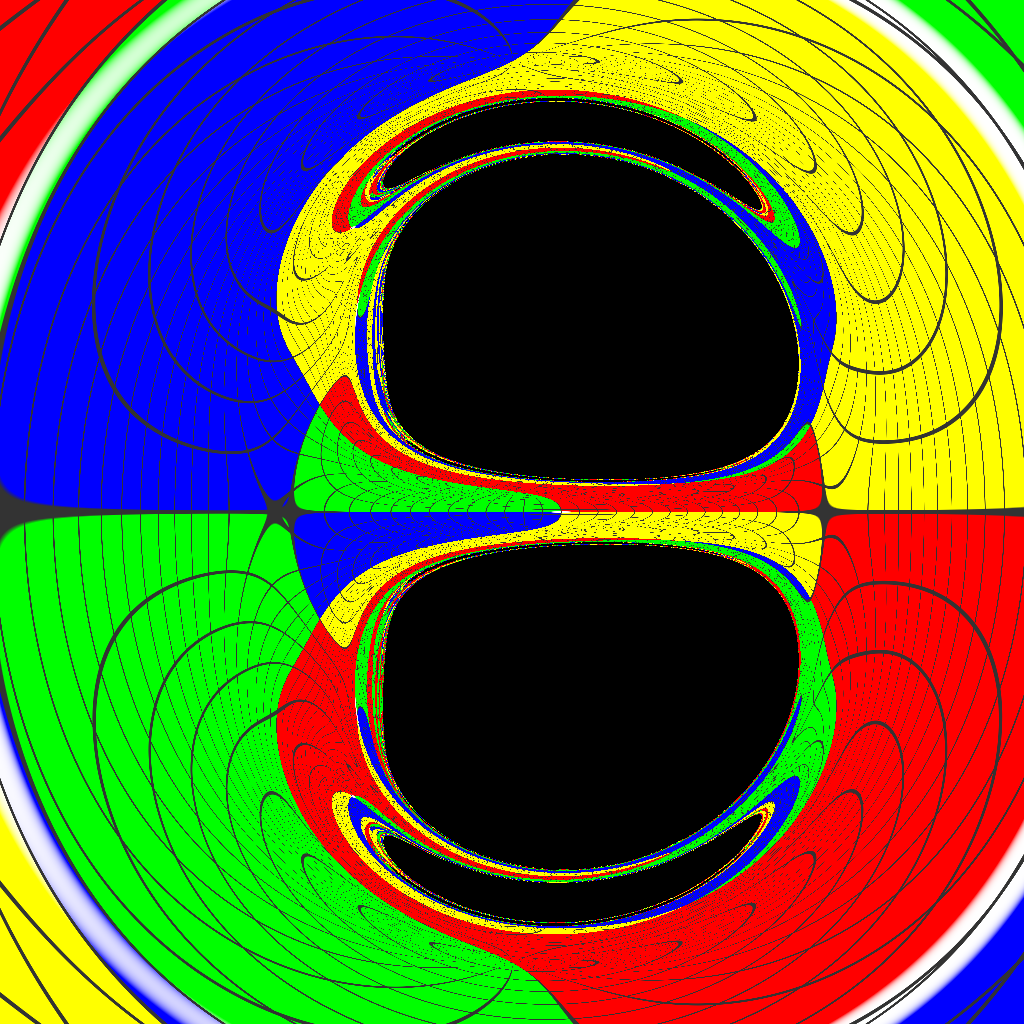}
	\includegraphics[scale=0.22]{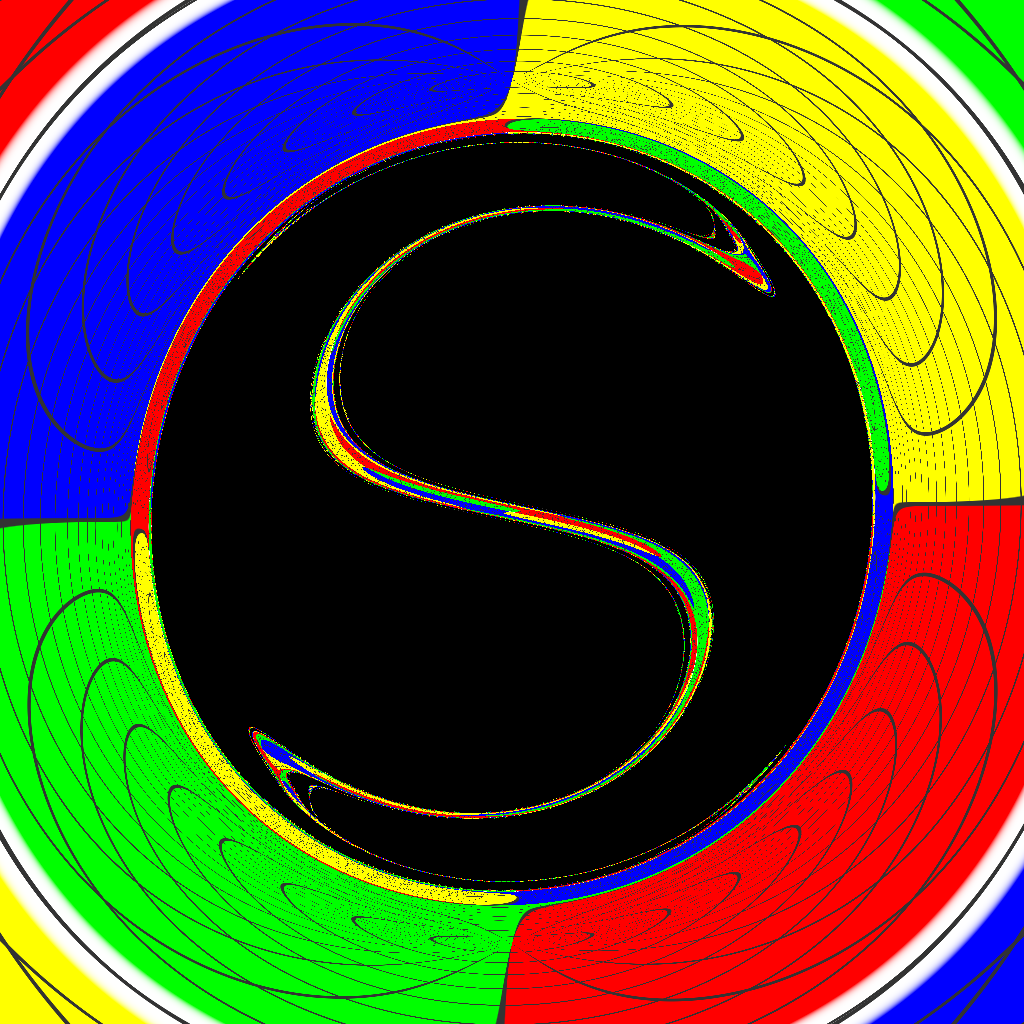}
	\includegraphics[scale=0.22]{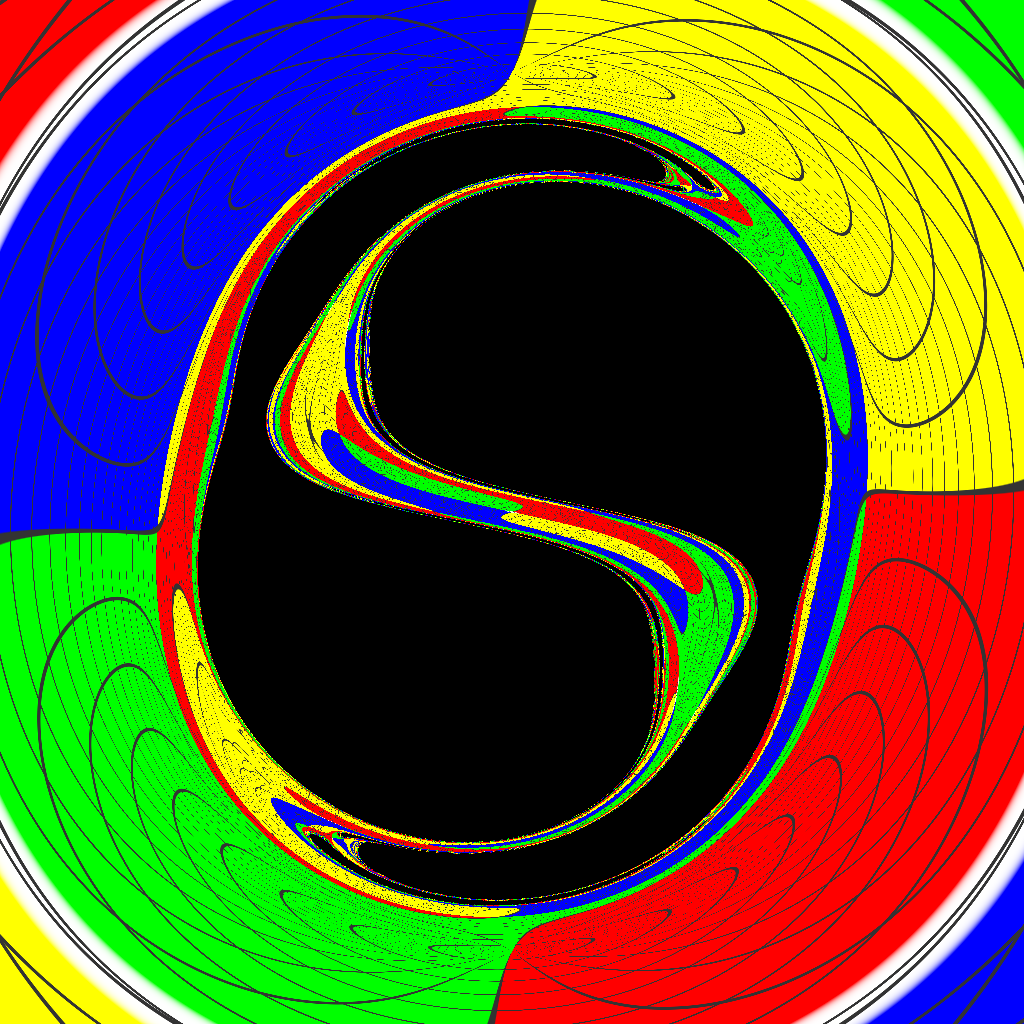}
	\includegraphics[scale=0.22]{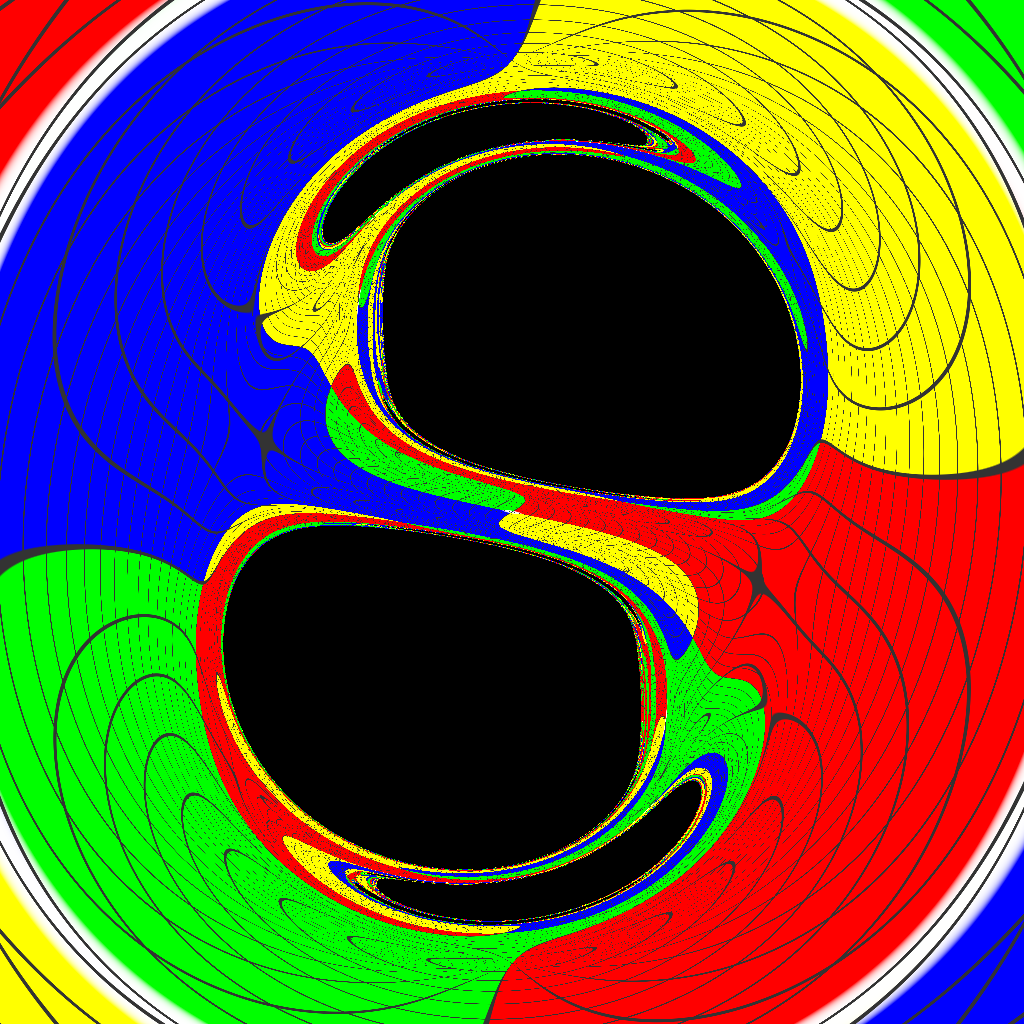}
	\caption{Shadow and gravitational lensing of the rapdily rotating ($J\approx J_{\text{e}}$) TW solution, starting with equal-mass MP solution and increasing the coordinate distance parameter $a$ for three different angular momentum configurations, namely the cases $(i),\ (ii)$ and $(iii)$ discussed in Sec.~\ref{subsecIVB}. In all the images, the observer is
		positioned at the equatorial plane ($z=0$) and at the radius $\rho = 15M$.}
	\label{Shadow}
\end{figure*}

We also computed some FPOs of the TW spacetime, as shown in Fig.~\ref{TW_FPOs}. Black points represent the BHs, with light trajectories shown in both colored and black lines. Dotted lines indicate LRs derived from $\mathcal{H}_+^{\text{TW}}$, while dashed lines correspond to LRs from $\mathcal{H}_-^{\text{TW}}$. Each image serves as a natural generalization of Fig.~\ref{MPFPOs}, illustrating how photon trajectories are influenced by different configurations of BH angular momentum. Each colored FPO in Fig.~\ref{MPFPOs} assumes some sort of 3 dimensional rotated version in Fig.~\ref{TW_FPOs}. The external pink orbit ``sees" the binary as almost a single gravitational center, whereas the internal purple and blue orbits see each center individually. 

In configuration $(i)$, with $J_1 = 0$ (top left), the blue trajectory is asymmetric with respect to the plane $z = 0$. In this setup, light rays are strongly dragged near the top BH, which carries angular momentum $J_2/M^2 = 0.1$, compared to the bottom BH with $J_1 = 0$. As a result, rays above $z = 0$ are more widely spaced, while trajectories below $z = 0$ are more closely clustered. 

In contrast, cases $(ii)$ and $(iii)$ exhibit blue light rays that are more symmetric with respect to the equatorial plane, due to the spacetime's reflection symmetry across this plane. Additionally, in case $(iii)$, involving oppositely rotating BHs, the LRs of $\mathcal{H}_+^{\text{TW}}$ always share the same $\rho$ coordinate with a corresponding LR of $\mathcal{H}_-^{\text{TW}}$, which is symmetrically reflected across the plane $z = 0$. This is a special feature of $odd$ $\mathbb{Z}_2$ symmetric spacetimes, which was discussed in more detail in Ref.~\cite{Moreira:2024sjq}.

It is worth noting that in configuration $(iii)$, although the purple and pink trajectories appear to be confined to a vertical two-dimensional plane, this is not the case. The opposite angular momenta prevent these trajectories from developing the three-dimensional structure observed in configurations $(i)$ and $(ii)$. However, there are frame dragging effects that add to the trajectories a wiggly appearance, which deviates from the behavior obtained in Fig.~\ref{MPFPOs}.

\section{Shadow and gravitational lensing}
\label{Sec5}

Here, we analyze the shadow and gravitational lensing effects associated with the TW solution. The possibility to analytically trace the edge of a BH shadow depends on whether the photon motion equations are Liouville-integrable, which does not seem to be the case of the metric~\eqref{TW1}, for which no Carter constant is known; alongside the other conserved quantities like energy, angular momentum, and mass.

In spacetimes where geodesic equations are not integrable, variable separation is not possible, and photon trajectories may exhibit chaotic behavior. In such scenarios, analytical methods for determining the BH shadow become less powerful. Instead, numerical techniques like backward ray-tracing are very useful. This method involves tracing null geodesics backward from the observer’s location to determine whether they are captured by the BH or escape to a distant celestial sphere. 

Consequently, we rely on numerical integration of the geodesic equations for massless particles. The initial conditions for this integration are derived by projecting the photon's momentum onto the observer’s tetrad frame—details of which can be found in Ref.~\cite{da2015black}. The initial conditions for a ZAMO frame are given by
\begin{equation}\label{IC1}
	E=\frac{(H_+H_-)^{1/4}}{\sqrt{\rho^2 H_+H_--{\omega_\varphi^0}^2}}\left(\rho+\frac{\omega_\varphi^0\cos\alpha\sin\beta}{\sqrt{H_+H_-}}\right),
\end{equation}

\begin{equation}\label{IC2}
	p_\rho= (H_+H_-)^{1/4}\cos \alpha \cos\beta,
\end{equation}

\begin{equation}\label{IC3}
	p_z= (H_+H_-)^{1/4}\sin\alpha,
\end{equation}

\begin{equation}\label{IC4}
	L=\frac{\sqrt{\rho^2 H_+H_--{\omega_\varphi^0}^2}}{(H_+H_-)^{1/4}}\cos \alpha \sin\beta,
\end{equation}
where $\alpha,\ \beta$ are the observation angles and Eqs.~\eqref{IC1}-\eqref{IC4} must be
evaluated at the observer coordinates.

The initial conditions for the photons are determined by the observation angles. Using the PyHole Python package~\cite{Cunha:2016bjh}, we simulated 1024×1024 light ray trajectories by varying these angles. Each null geodesic contributes a single pixel to the lensed image: geodesics that are captured by the event horizon result in black pixels, while those that escape are assigned colors. The celestial sphere was divided into four quadrants, each one characterized by a different color: red, green, blue, and yellow. Additionally, a white circular patch was placed at the center of the celestial sphere, directly in front of the observer.

\subsection{Equatorial plane images}

We display in Fig.~\ref{MP_Shadow} the case $J/M^2 = 0$, which corresponds to the MP subcase with $a/M=1$, to provide a consistent reference. Althought the shadows of MP and double-Schwarzschild are significantly similar (see Figure 2 of Ref.~\cite{Cunha:2018gql}), the MP image does not exhibit the discontinuity on the lensing as it does for the double-Schwarzschild, since there are no conical singularities.   

Figure~\ref{Shadow} presents the shadows and gravitational lensing of the TW solution for an observer located on the equatorial plane ($z = 0$) at a radial coordinate of $\rho = 15M$. The three rows correspond to the angular momentum configurations $(i)$, $(ii)$, and $(iii)$ described in~\ref{subsecIVB}, respectively. To calculate the images we fix the magnitude of the angular momentum to $J/M^2 = 0.499\approx J_{\text{e}}/M^2$. For each configuration, the coordinate distance parameter $a$ takes three values: $a/M = 0.5$, $1.0$, and $1.5$, with the distance increasing from left to right along each row.


The shadows reveal no significantly novel characteristics and it seems hard to distinguish them from the double-Kerr~\cite{Cunha:2018cof}. In particular, the characteristic 'eyebrows'~\cite{Shipley:2016omi, Yumoto:2012kz, Nitta:2011in}, located above and below each primary shadow, are a typical feature of double-BH configurations. 

\begin{figure}[t]
	\centering
	\includegraphics[width=\columnwidth]{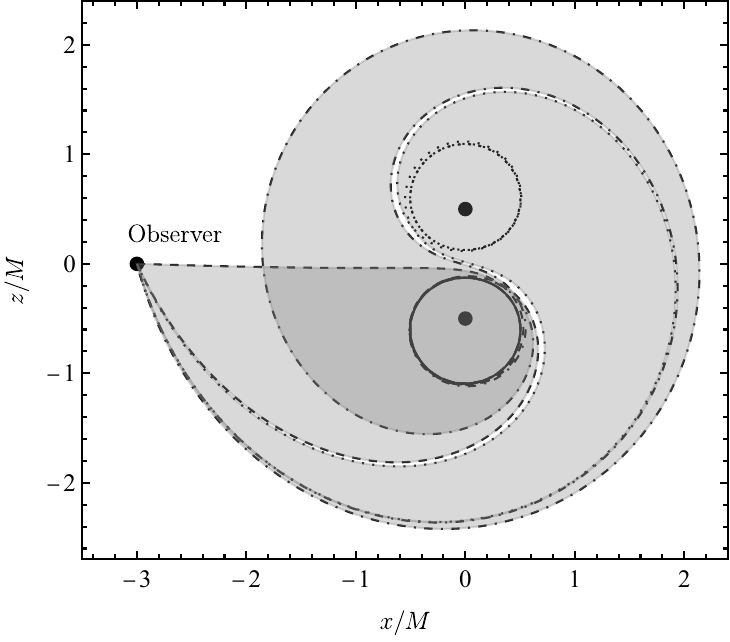}
	\includegraphics[width=\columnwidth]{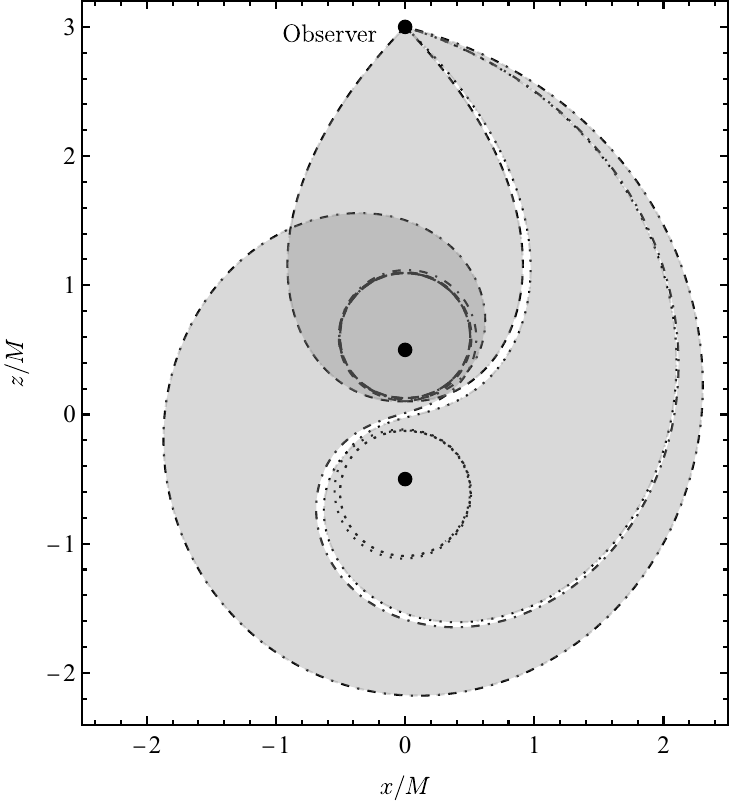}
	\caption{Light trajectories corresponding to the eyebrow features of the MP's shadow with two BHs. In the top panel, the observer is located at the equator, while in the bottom panel, the observer is positioned above the BHs.}
	\label{Eyebrow}
\end{figure}

These features appear in simulations as secondary shadows, which are more clearly visualized in Fig.~\ref{Eyebrow}. For illustration, the top panel shows an equatorial observer facing equal-mass MP BHs, with $a/M = 0.5$. We display a vertical cross-section of light rays corresponding to each shadow seen in Fig.~\ref{Shadow}. The group of geodesics that either fall into the bottom BH or end on the FPO surrounding it corresponds to the primary shadow of the bottom BH. This group is highlighted in gray regions in the top panel of Fig.~\ref{Eyebrow}, with boundaries marked by black dashed lines. Similarly, geodesics bounded by the dotted line either fall into the top BH or become trapped by its nearby FPO, forming a secondary shadow below the primary shadow. This secondary shadow, or the bottom eyebrow, is much thinner due to the smaller viewing angle of these geodesics. A similar and even thinner shadow layer, associated with the dot dashed line, exists below this bottom eyebrow, associated with the bottom BH. This pattern repeats infinitely, creating a fractal-like structure of nested shadows.

The first row, which represents case $(i)$, the bottom BH has $J=0$. The eyebrows become significantly larger when the BHs possess angular momentum. Hence, its shadow looks distorted only by frame dragging effects of the top BH and is not, as expected, $\mathbb{Z}_2$ symmetric.

In contrast with case $(i)$, the cases $(ii)$ and $(iii)$ have shadows that are even and odd $\mathbb{Z}_2$ symmetric, respectively~\cite{Moreira:2024sjq}. We additionally remark that the rotation effects on the shadows are more noticeable in case $(iii)$, as the eyebrows merge with the primary shadows.


\subsection{Off-equatorial plane images}

\begin{figure*}
	\centering
	\includegraphics[scale=0.22]{TW_Spherical_M1eM2e1_ae1d0_J1e0_J2e0.5}
	\includegraphics[scale=0.22]{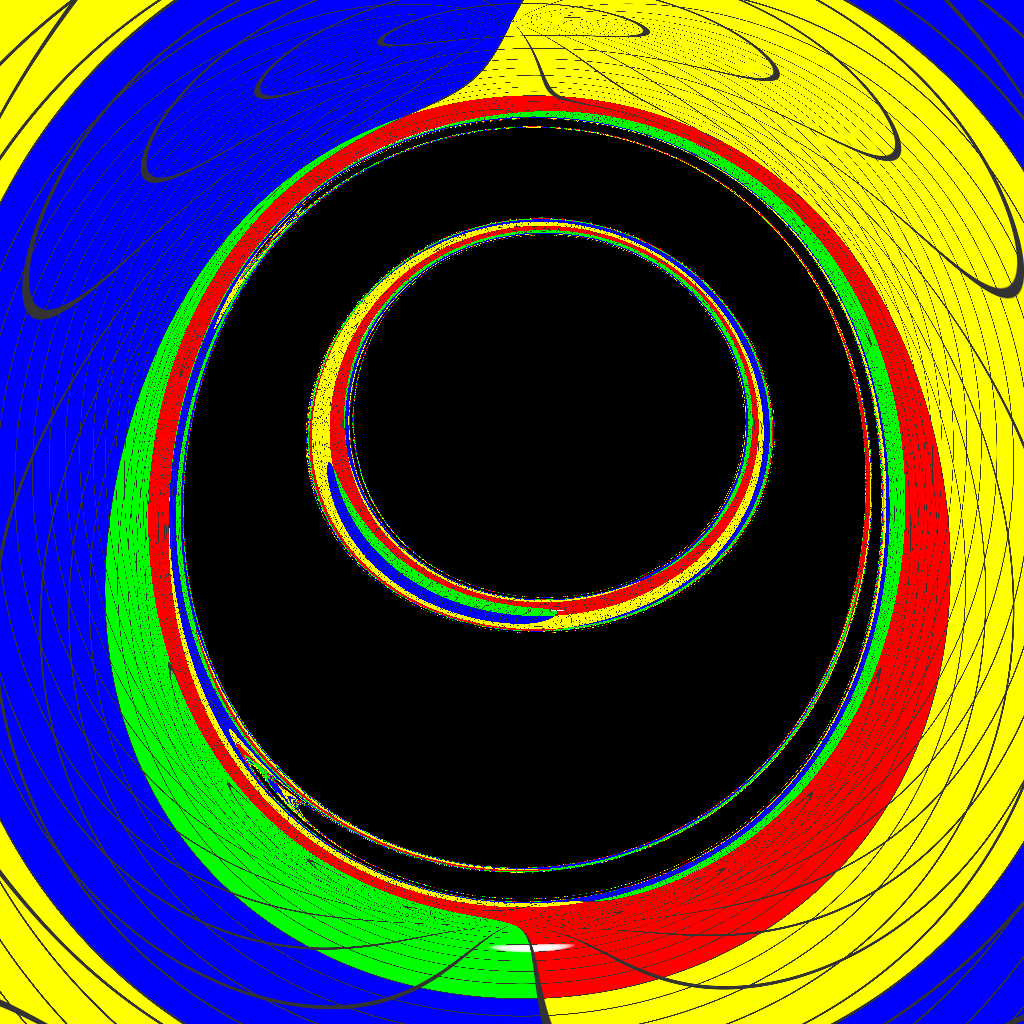}
	\includegraphics[scale=0.22]{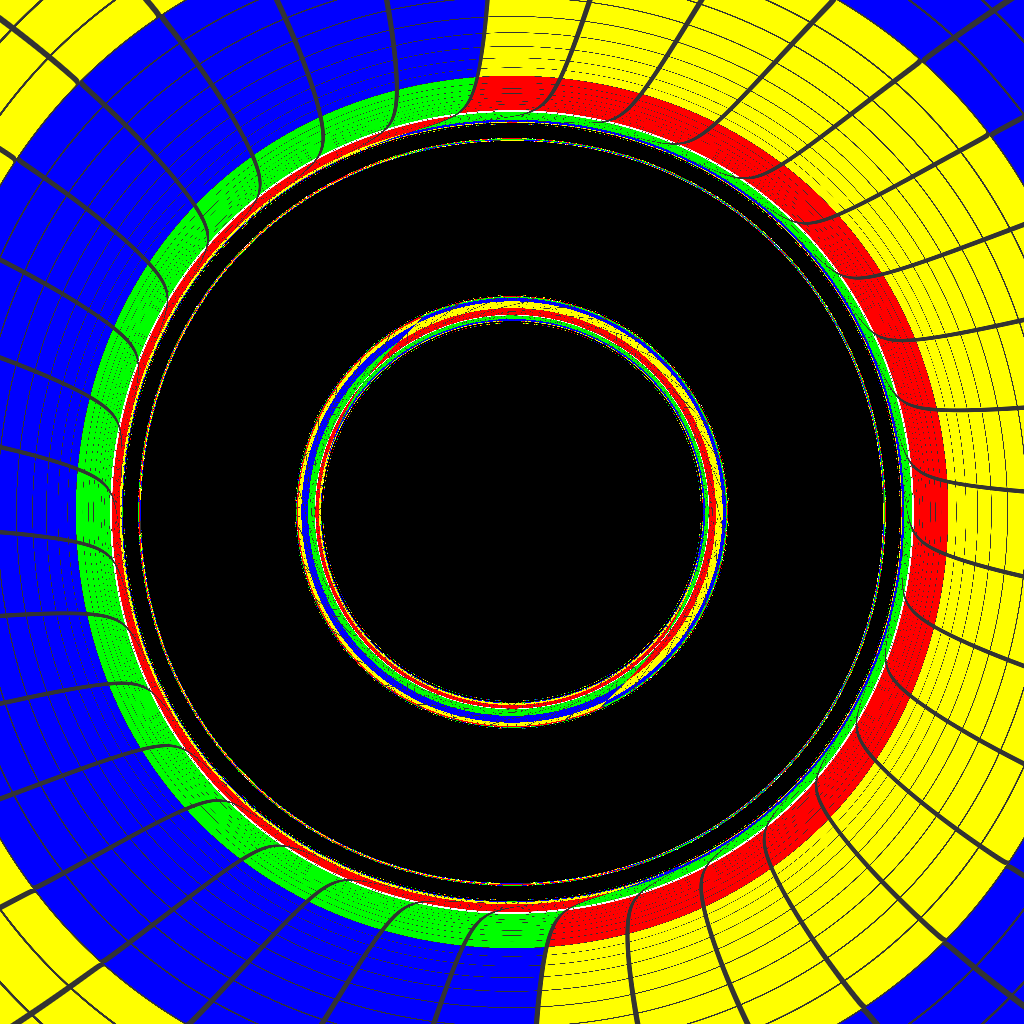}
	\includegraphics[scale=0.22]{TW_Spherical_M1eM2e1_ae1d0_J1eJ2e0.5}
	\includegraphics[scale=0.22]{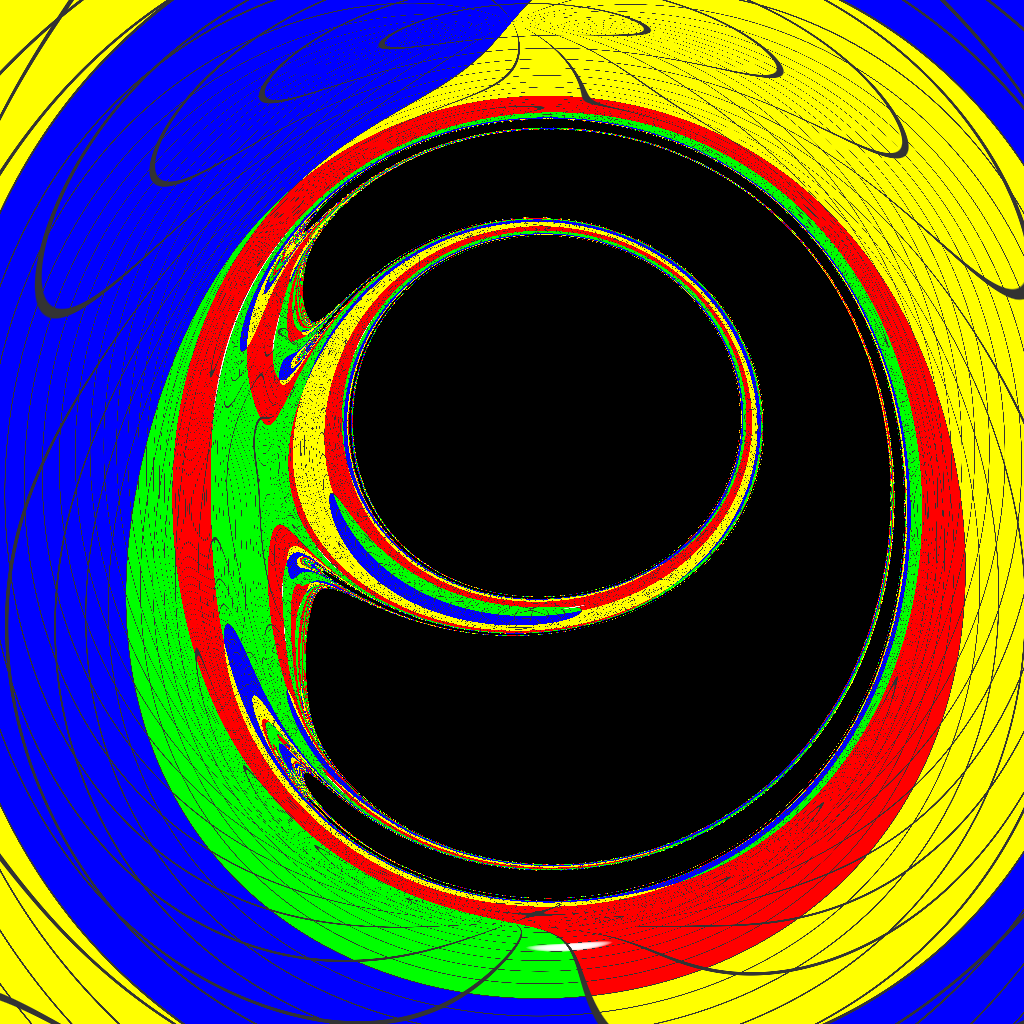}
	\includegraphics[scale=0.22]{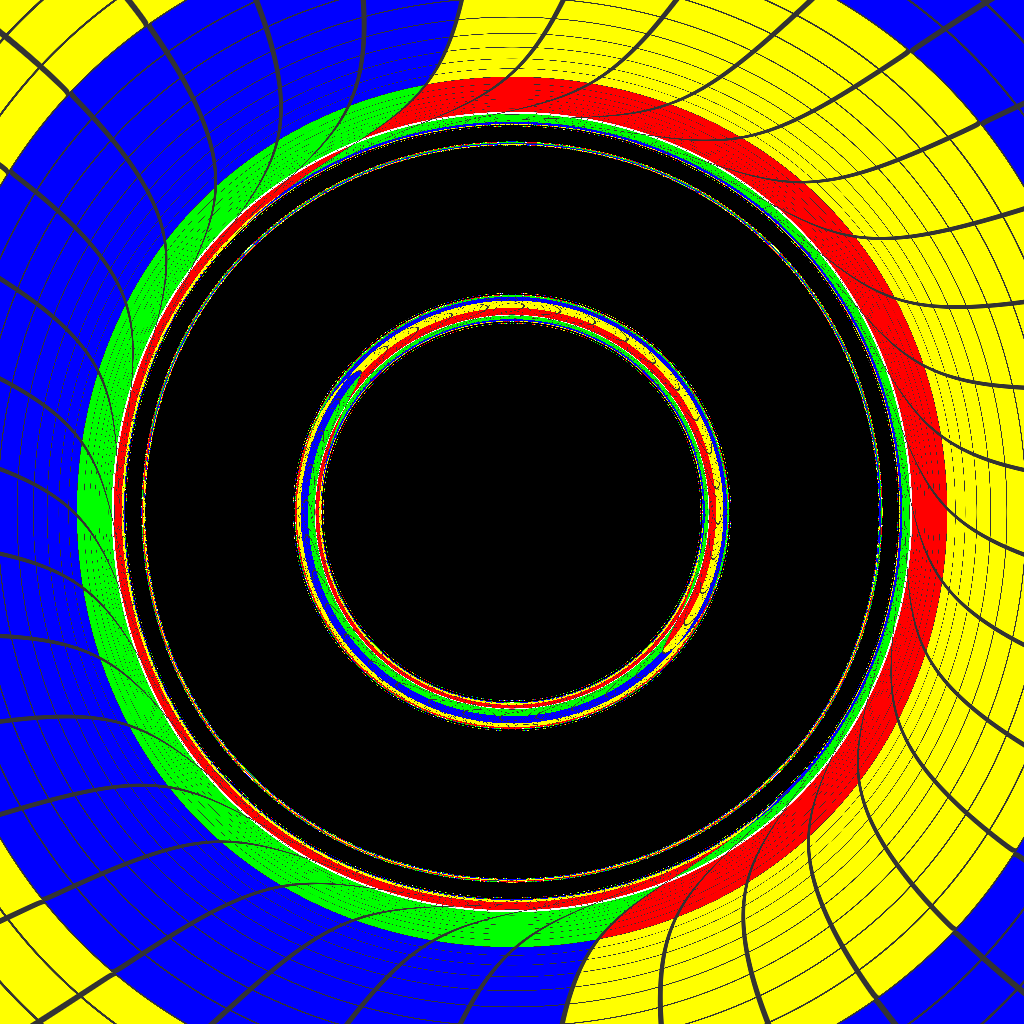}
	\includegraphics[scale=0.22]{TW_Spherical_M1eM2e1_ae1d0_mJ1eJ2e0.5}
	\includegraphics[scale=0.22]{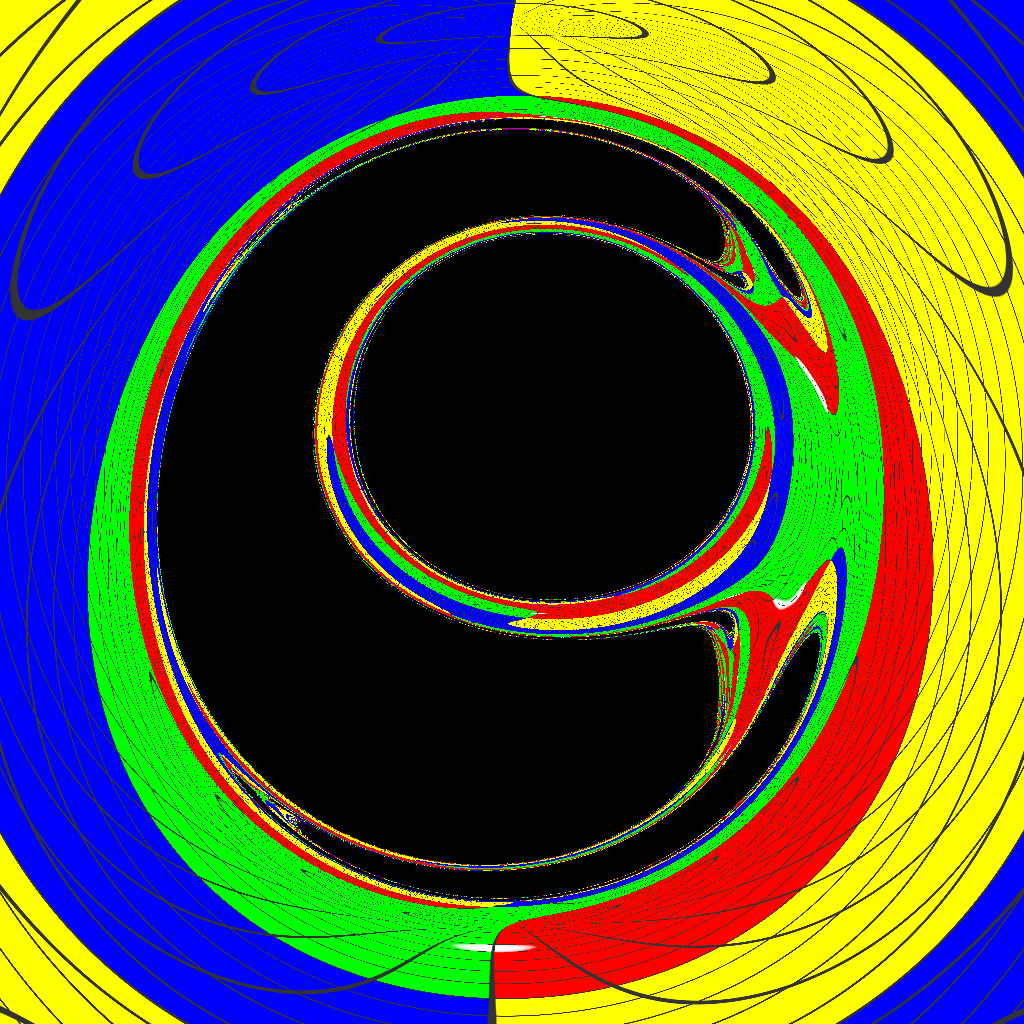}
	\includegraphics[scale=0.22]{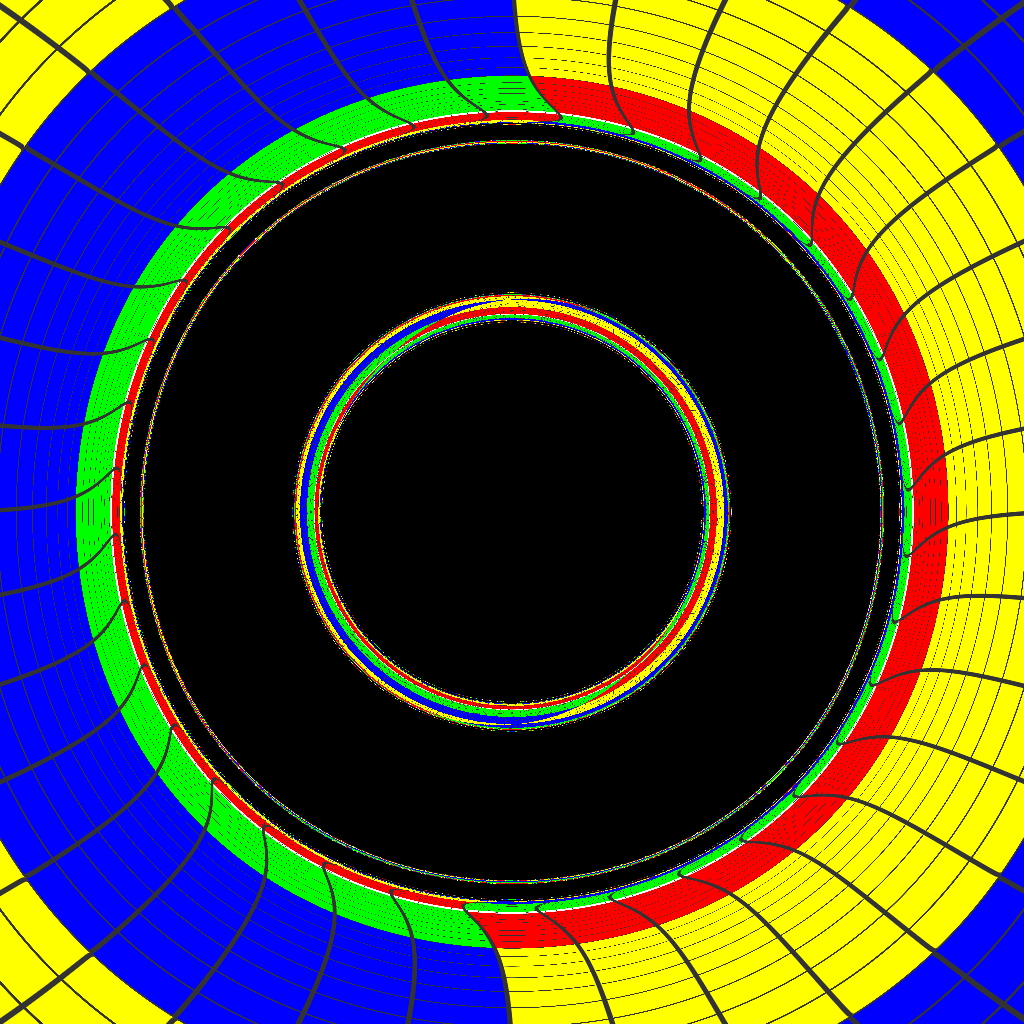}
	\caption{Shadow and gravitational lensing of the TW solution, starting with an observer at the equator for near-extremal $J \approx J_{\text{e}}$ on the left of each row, corresponding to configurations $(i)$, $(ii)$, and $(iii)$, respectively. Moving from left to right, the observer's position changes with increments of $\theta = \pi/4$ and ending at the top of the binary.}
	\label{ShadowOff}
\end{figure*}

Next, we examine the shadow and gravitational lensing when the observer is positioned off the equatorial plane. For this analysis, we consider the near-maximum angular momentum $J\approx J_{\text{e}}$ configuration for the same cases $(i)$, $(ii)$, and $(iii)$ on each row. The observer’s radial distance remains fixed, but the polar angle $\theta \equiv \arctan\frac{\rho}{z}$ is varied, starting from $\theta = 0$ and ending at $\theta = \pi/2$. We use equal step increments of $\delta \theta = \pi/4$ in the observer's $\theta$ coordinate.

The simulated images are shown in Fig.~\ref{ShadowOff}. The leftmost image on each row corresponds to an observer positioned at the equator, as in previous analyses, for reference. The rightmost image shows the perspective from directly above the two BHs. Due to axial symmetry, the shadow boundary for $\theta=0$ observer's forms a perfect circle at the center, representing the shadow of the top BH. The central portion of the shadow corresponds to geodesics within the gray region in Fig.~\ref{Eyebrow}, bounded by the dashed line, which are captured by the top BH. Surrounding this core shadow, there is a ring-shaped shadow created by the bottom BH, followed by a thinner outer ring associated with a secondary shadow of the top BH. These two ring-like shadows are related with ther dotted and dot dashed lines in Fig.~\ref{Eyebrow}, respectively. This nested structure of shadows and rings continues infinitely, reflecting a similar repeating pattern seen in eyebrow formation.

\section{Final remarks}
\label{Sec6}

The TW spacetime~\cite{Teo:2023wfd} represents a novel class of exact solutions obtained in the context of KK theory, as a multi-center generalization of the RL BH~\cite{Rasheed:1995zv,Larsen:1999pp}. The TW solution is also a regular rotating generalization of MP solution, where the equilibrium established is ensured not only by the electromagnetic field, but also by the dilaton field. As is Refs.~\cite{Herdeiro:2023mpt,Herdeiro:2023roz}, the presence of a scalar field is fundamental to achieve a proper equilibrium. 

Our analysis of LRs focused on using the TC formalism to examine the various LR configurations across different parameter choices. We revisited the properties of the equal-mass MP spacetime through the lens of the TC method, exploring all possibilities within the parameter space. This leads to three distinct LR configurations: a compact setup with two equatorial LRs for $a < a_i$, a transient regime with four LRs—two of which are equatorial—for $a_i < a < a_f$, and a spread-out BH regime for $a > a_f$, featuring two LRs near each gravitating center. The transition between the regimes $a < a_i$ and $a > a_f$ is illustrated in Fig.~\ref{LRcoal}, aiding in a clearer understanding of the process. Additionally, the stability of the LRs has been analyzed and summarized in Tab.~\ref{LRtable}.

In the rotating case, with five parameters, the LR structure becomes significantly richer than in the static scenario. The angular momentum splits each LR of the static case into two, one for each sense of rotation, resulting in a maximum of 8 LRs. However, LRs with opposite TC can merge and annihilate, leading to configurations with 6 or 4 LRs. Whether there are 4, 6, or 8 LRs depends on the choice of solution parameters. Focusing on equal-mass BHs and three specific angular momentum configurations, namely: $(i): J_1 = 0, J_2 = J$; $(ii): J_1 = J_2 = J$; and $(iii): -J_1 = J_2 = J$, we mapped the parameter space and identified all possible LR configurations (see Fig.~\ref{LR_Jcrit}).

We also calculated some FPOs for the TW spacetime and compared them with those previously reported for the MP spacetime. As before, we focused on the specific angular momentum configurations $(i)$, $(ii)$, and $(iii)$. The MP orbits shown in Fig.~\ref{MPFPOs}, initially confined to a vertical plane, become frame-dragged when angular momentum is introduced, transforming into the orbits displayed in Fig.~\ref{TW_FPOs}. Notably, the pink and purple orbits in Fig.~\ref{TW_FPOs}, for oppositely rotating BHs appear to lie within a vertical plane; however, due to the dragging effect, this is not the case. As these orbits evolve, they experience dragging in opposite directions, ultimately forming a 1-dimensional submanifold.

We also investigated the shadows and gravitational lensing of the TW solution. In nonintegrable spacetimes, as the TW binary seems to be, the geodesic equations cannot be separated, making analytical shadow calculations not possible in a general context. Instead, shadows can be determined through numerical simulations using backward ray-tracing techniques. The shadow's 'eyebrows' grow significantly when the BHs have angular momentum, and its shape adopts a D-like appearance, similar to that of the double-Kerr spacetime. While the TW and double-Kerr spacetimes may differ in many respects, they share similar shadow characteristics.

We remark that the TW spacetime reported in Ref.~\cite{Teo:2023wfd} chooses the BHs to have equal electric and magnetic charges, which corresponds to the black line displayed in Fig.~\ref{Extremal}, but this is not mandatory.  We chose to exhibit the extremal surface of the RL BH to show how much more can be explored. All our analyses are valid just for this  black line segment in Fig.~\ref{Extremal}. It is an interesting task to generalize the TW spacetime, and the analysis in this paper, for arbitrary charge values abiding the extremality constraint.

It has been reported that a quasi-static approach to simulate images for stationary BBHs can serve as a good approximation for a fully dynamical binary system~\cite{Cunha:2018cof}. Although this method was successfully applied to the double-Schwarzschild case, no implementation for the double-Kerr configuration has been made, due to the complexity of the metric.  Given that the TW spacetime exhibits a similar shadow pattern to the double-Kerr but has a simpler metric, it is more likely that the quasi-static method can be effectively applied to this case, offering a useful approximation for the images of a fully dynamical Kerr binary.

\begin{acknowledgments}
	We are grateful to Funda\c{c}\~ao Amaz\^onia de Amparo a Estudos e Pesquisas (FAPESPA), Conselho Nacional de Desenvolvimento Cient\'ifico e Tecnol\'ogico (CNPq) and Coordena\c{c}\~ao de Aperfei\c{c}oamento de Pessoal de N\'ivel Superior (CAPES) -- Finance Code 001, from Brazil, for partial financial support. 
ZM and LC thank the University of Aveiro, in Portugal, for the kind hospitality during the completion of this work.
This work is supported by the Center for Research and Development in Mathematics and Applications (CIDMA) through the Portuguese Foundation for Science and Technology (FCT -- Fundaç\~ao para a Ci\^encia e a Tecnologia) under the Multi-Annual Financing Program for R\&D Units, PTDC/FIS-AST/3041/2020 (\url{http://doi.org/10.54499/PTDC/FIS-AST/3041/2020}),  2022.04560.PTDC (\url{https://doi.org/10.54499/2022.04560.PTDC}) and 2024.05617.CERN (\url{https://doi.org/10.54499/2024.05617.CERN}). This work has further been supported by the European Horizon Europe staff exchange (SE) programme HORIZON-MSCA-2021-SE-01 Grant No.\ NewFunFiCO-101086251.
\end{acknowledgments}


\bibliography{ref}

\end{document}